\documentclass[a4paper,11pt]{article}
\pdfoutput=1 % if your are submitting a pdflatex (i.e. if you have
\usepackage{jcappub} % for details on the use of the package, please
\usepackage[T1]{fontenc} % if needed

\usepackage{subcaption}
\usepackage{xcolor}

\title{Primordial magnetic fields: consistent initial conditions and impact on high-z structures}

\author[a,b]{Pranjal Ralegankar,}
\author[a,b]{Mak Pavi\v{c}evi\'{c},}
\author[a,b,c,d, e]{and Matteo Viel}

\affiliation[a]{SISSA - International School for Advanced Studies, Via Bonomea 265, 34136 Trieste, Italy}
\affiliation[b]{INFN – National Institute for Nuclear Physics, Via Valerio 2, I-34127 Trieste, Italy}
\affiliation[c]{IFPU, Institute for Fundamental Physics of the Universe, via Beirut 2, 34151 Trieste, Italy}
\affiliation[d]{INAF, Osservatorio Astronomico di Trieste, Via G. B. Tiepolo 11, I-34131 Trieste, Italy}
\affiliation[e]{ICSC — Italian Research Center on High Performance Computing, Big Data
and Quantum Computing, Via Magnanelli 2, 40033 Casalecchio di Reno (BO), Italy}

\emailAdd{pralegan@sissa.it}
\emailAdd{mpavicev@sissa.it}
\emailAdd{viel@sissa.it}

\abstract{Primordial magnetic fields (PMFs) can enhance matter power spectrum on small scales ($\lesssim$ Mpc) and still agree with bounds from cosmic microwave background (CMB) and Faraday rotation measurements. As modes on scales smaller than Mpc have already become non-linear today, exploring PMFs' impact on small-scale structures requires dedicated cosmological simulations. Here, for the first time, we perform a suite of hydrodynamical simulations that take into account the different impacts of PMFs on baryons and dark matter. Specifically, in the initial conditions we displace particles according to the Lorentz force from PMFs. We also highlight the large theoretical uncertainty in the peak enhancement of the matter power spectrum due to PMFs, which was not considered in previous studies. We present halo mass functions and show that they can be accurately reproduced using Sheth-Tormen formalism. Moreover, we show that PMFs can generate galaxies with baryon fraction several times larger than the cosmic average at high redshifts. This is simply a consequence of the fact that PMFs enhance baryon perturbations, causing them to be larger than dark matter perturbations. We argue that this scenario could be tested soon by obtaining accurate estimates of the baryon fraction in high redshift galaxies.}

\begin{document}
	\maketitle
	\flushbottom
	
	\section{Introduction}
	\label{sec:intro}
	
	Most of the observed magnetic fields in the cosmos are attributed to astrophysical sources, such as those generated by stars or by the motion of ionised particles within galaxies. However, there have been hints of magnetic fields even in the depths of cosmic voids, where traditional magnetic sources are mainly absent \cite{doi:10.1126/science.1184192, HESS:2014kkl, Finke:2015ona, VERITAS:2017gkr, AlvesBatista:2021sln}. These void magnetic fields could instead trace back to processes occurring at the very birth of the universe, such as inflation or phase transitions \cite{Vachaspati:2020blt, Subramanian:2015lua}. Thus, if the presence of primordial magnetic fields (PMFs) is confirmed, it would provide valuable insights into new fundamental physics.
	
	A distinctive signal for PMFs can come from their impact on small-scale structures ($M<10^{12}\ {\rm M}_{\odot}$). Specifically, the Lorentz force from the PMFs on the baryon plasma leads to the growth of baryon inhomogeneities \cite{wasserman97, Kim:1994zh, Subramanian:1997gi}, with larger growth on smaller scales. Current observations of the large-scale structure are in agreement with  $\Lambda$CDM cosmology and hence PMFs that enhance matter power spectrum on scales larger than $\sim 1$ Mpc are ruled out or tightly constrained \cite{Shaw:2010ea}. This provides an upper limit on PMFs averaged over 1 Mpc, $B_{\rm 1 Mpc}\lesssim$ 1 nG. Observations from the cosmic microwave background and Faraday rotation measurements also provide similar constraints and limit $B_{\rm 1 Mpc}\lesssim$ 1 nG \cite{Planck:2015zrl, PhysRevLett.116.191302}. To probe weaker magnetic fields, we need to look at matter distribution on scales smaller than Mpc, where density perturbations have already become non-linear. Thus, to obtain robust constraints on PMFs from small-scale structures ($M<10^{12}\ {\rm M}_{\odot}$), it is imperative to carefully study how PMFs affect structure formation processes.
	
	Several studies have previously attempted to constrain PMFs by investigating their influence on
	small-scale structures, including their effects on reionization \cite{Katz:2021, Tashiro:2006, Sethi:2004pe, pandeysethi:2015}, the 21 cm signal \cite{Cruz:2023rmo, Schleicher:2008hc}, the Lyman-$\alpha$ forest \cite{Chong13,khan13,montanino2017}, properties of dwarf galaxies \cite{Sanati_2020}, or line intensity mapping experiments \cite{Adi23}.\footnote{Alternatively, several other works have studied the evolution of primordial magnetic fields on the scales of galaxy clusters, but without including their boost on the small scale power spectrum; see for example \cite{vazza21b,vazza21,Mtchedlidze:2021bfy}.} However, these studies assume that structure formation with PMFs occurs similarly to $\Lambda$CDM cosmology, except with an enhanced small-scale matter power spectrum. While this approach accurately estimates halo abundance, it could significantly misjudge the properties of galaxies residing in those halos. Moreover, these studies do not consider the large theoretical uncertainty in the peak enhancement of the matter power spectrum due to PMFs. Thus, they might have overestimated the constraints on PMFs from reionization, Lyman-$\alpha$ forest, 21cm intensity mapping or other tracers of structure formation processes.

	%Specifically, the studies that performed structure formation simulations generated initial conditions for matter particles by only using the total matter power spectrum \cite{Katz:2021,Sanati:2020}.
	
	%Typically these investigations rely on either a semi-analytical treatment of the most important physical processes or on N-body simulations in which the extra power in the baryonic fluid is taken into account by a implementing a given total (linear) matter power in the initial conditions and assign it to the dark matter. It is clear that this implementation is just an approximation and fails in capturing the richer implications of a full hydrodynamical simulations.
	
	In this study, we carefully outline the theory behind PMF's influence on matter perturbations and take the first step to perform cosmological simulations that are consistent with the theory. Specifically, we modify the initial conditions of cosmological simulations by displacing particles according to the Lorentz force from PMFs. Such a procedure takes into account that PMFs induce vortical (solenoidal) motion in baryons as well as that density perturbations sourced by PMFs are non-gaussian. More importantly, it considers that PMFs only enhance baryon density perturbations, while the dark matter perturbations are enhanced indirectly via the gravitational influence of baryon perturbations. We follow the evolution of gas and dark matter particles by performing hydrodynamical simulations, albeit without magnetohydrodynamics (MHD). We begin our simulations after the time when the gravitational force is expected to be stronger than the Lorentz force from PMFs, and hence the neglection of MHD in our simulation is not expected to have a large impact on halo formation.
	
	%Initial conditions generated by the total matter power spectrum cannot take into account the differential impact of PMFs on baryon and dark matter. For instance, on scales where PMFs enhance the matter power spectrum, baryon density perturbations are larger than dark matter density perturbations,  Additionally,  Finally, the positions of baryon particles and PMFs are correlated and this correlation can have a non-trivial impact as matter particles collapse to form halos.

	%In this study, we take the first step to explore the impact of PMFs on structure formation that goes beyond the small-scale enhancement of power, by implementing, for the first time, the actual power spectra of the two different components in the initial conditions and by following them down to low redshift.
	%Specifically, we take into account the fact that PMFs only enhance baryon density perturbations. Thus, the dark matter will eventually fall into the gravitational potential wells created by baryons first, which is in sharp contrast to the current paradigm where baryons fall into the dark matter potential wells. 
	To highlight the importance of improved simulations with PMFs, in this study, we primarily focus on PMFs' impact on the baryon fraction of high-redshift galaxies. As baryon fraction in structures is largely determined by initial conditions and the subsequent evolution under gravity, our results should remain robust even after including MHD affects. Furthermore, a focus on high-redshift also reduces astrophysical uncertainties. 
	%Consequently, have a larger baryon fraction in the early universe compared to the cosmic average. A larger baryon fraction implies a larger availability of gas to form stars, and hence a higher star formation rate.
	
	Notably, recent JWST observations 
	%have observed high redshift galaxies with very high stellar masses and with substantial star formation rates 
	have reported the observation of high-redshift galaxies with large stellar masses and substantial star formation rates
	\cite{labbe23,robertson23, 2023arXiv230902492X, tacchella23, bunker23, maiolino23, chemerynska23}, providing measurements of their luminosity function \cite{donnan23} and other physical properties. These findings have opened up the possibility that it might be difficult to reconcile the cosmological model with observational properties of high-redshift galaxies and, for example, non standard dark energy models might reconcile the tension \cite{menci22} or dark matter models can be constrained \cite{maioviel}. However, such results are under debate and while some cosmological galaxy formation simulations broadly succeed in reproducing the population of galaxies \cite{kannan22,garaldi23,mccaffrey23,prada23}, it is still possible that new fundamental physics or still poorly understood astrophysical processes like radiation driven feedback or dust might be at play \cite{li23,ferrara23}. In this work, we do not aim at providing a realistic physical description of high-redshift galaxies but we are more interested in discussing the relative differences between a scenario in which PMFs are present with respect to the standard cosmological model.
	
	%To quantitatively explore how PMFs affect the abundance of high stellar mass galaxies in the early universe, we perform hydrodynamical simulations, albeit without magnetohydrodynamics (MHD). We account for PMFs by appropriately generating initial conditions for baryons and dark matter according to the Lorentz force exerted by PMFs. We focus our simulations on length scales where the gravitational force is expected to be stronger than the Lorentz force from PMFs, and hence the neglection of MHD in our simulation is expected to have a minimal impact on halo formation.
	
	We find that the baryon fraction of halos that are sourced by realistic PMFs is about two times larger than the cosmic average at high redshifts, $z>4$. This is due to the fact that PMF-enhanced perturbations have dark matter density perturbations smaller than baryon perturbations.  A larger baryon fraction implies a larger availability of gas to form stars, and hence a higher star formation rate. 
	Furthermore, isocurvature between perturbations sourced from inflation and those from PMFs is found to increase the scatter in the baryon fraction. Other novel features such as non-gaussianities and vortical motion in baryons are found to be largely unimportant in the determination of baryon fraction.
	
	This paper goes beyond previous works that performed cosmological simulations with PMFs \cite{Sanati:2020, Katz:2021} by taking into account vorticity, non-gaussianities, the different impacts on baryons and dark matter, and the uncertainty in the peak of the matter power spectrum. However, the results in this work do not completely reflect the impact of PMFs as we do not perform full MHD simulations, which we leave to future work. Furthermore, we note that our formalism does not take into account in a self-consistent way the small-scale physics of star formation in the presence of magnetic fields. The inclusion of appropriate star-formation physics would require a dedicated investigation at very high resolution and is beyond the scope of this paper (see e.g. \cite{koh21}). 
	
	This paper is organized as follows. In section~\ref{sec:theory} we review the theory behind PMF's impact on cosmological density perturbations. We analytically show why high redshift halos are expected to have larger baryon fractions and also outline some simplifying assumptions in our analysis. Additionally, we highlight how the peak enhancement of the matter power spectrum has an uncertainty of orders of magnitude. In section~\ref{sec:sim}, we discuss the results of our hydrodynamical simulations whose initial conditions were modified to include the effects of PMFs. Finally, we conclude in section~\ref{sec:con}. 
	The details of many of our computations are relegated to appendices. In appendix~\ref{sec:AmbDiff}, we calculate the necessary ionization fraction needed for tight coupling of ions and neutrals. In appendix~\ref{sec:ngenic} we describe how we modified the {\scshape N-GenIC} code to produce initial conditions with PMF enhanced perturbations \footnote{https://www.h-its.org/2014/11/05/ngenic-code/} \cite{springel05}. In appendix~\ref{sec:pdf} we show the distribution of density perturbations that are sourced by PMFs. 
 %Finally, in appendix~\ref{sec:TestSimRuns}, we show that our Gadget simulations are performing as expected by rederiving the halo mass function known for standard cosmology.

	\section{Theoretical framework}\label{sec:theory}
	In this section, we review the evolution of cosmological density perturbations in the presence of PMFs. Specifically, we focus on the post-recombination universe where photons have decoupled from baryons. First, in section~\ref{sec:eqs}, we describe the magnetohydrodynamic (MHD) equations that govern the evolution of magnetic fields and baryons. Next, in section~\ref{sec:grow}, the evolution of dark matter and baryon density perturbations, under the influence of PMFs in the linear regime, is shown. We demonstrate that dark matter density perturbations are, for the most part, smaller than baryon density perturbations when PMFs source perturbations. This smaller value of dark matter perturbations is the primary cause behind large baryon fractions in high redshift halos.
	Finally, in section~\ref{sec:power}, we describe the power spectrum of matter perturbations sourced by PMFs. We highlight that the peak enhancement of the matter power spectrum is not theoretically certain and can change by orders of magnitude depending on the simplifying assumptions.

	\subsection{Governing MHD equations}
	\label{sec:eqs}
	After recombination, the fraction of ionised particles in the baryon plasma drops significantly from unity to order $\sim 10^{-4}$ (see, e.g., chapter 6 in \cite{PeeblesCosmology}). The PMFs induce motions in the ionised particles via the Lorentz force, and the ionised particles, in turn, drag the neutral particles with them via dipole interactions. Even though the fraction of charged particles is tiny, the interactions between the ions and neutral particles are strong enough to keep the whole baryon fluid tightly coupled \cite{Banerjee:2004df, Sethi:2004pe} (see appendix~\ref{sec:AmbDiff} for details).   
 Thus, the whole baryon fluid can be treated as a perfect conductor and magnetic fields evolve according to the ideal MHD equation,
	\begin{align}\label{eq:induction}
		\frac{\partial \vec{B}}{\partial t}&=\frac{1}{a}\nabla\times(\vec{v}_{\rm b}\times \vec{B}).
	\end{align}
	Here $\vec{B}$ is the comoving magnetic field, $\vec{B}=a^2\vec{B}_{\rm phys}$, with $\vec{B}_{\rm phys}$ being the physical magnetic field, $\vec{v}_{\rm b}$ is the baryon bulk velocity, $a$ is the scale factor, and $t$ is physical time.
	
	The motion of baryons is influenced by the Lorentz force from the PMFs. As we are primarily interested in scales that are within the horizon by the time of recombination ($l<$Mpc), the influence of PMFs on baryons is accurately captured by \cite{Kim:1994zh},
	\begin{align}
		\frac{\partial \vec{v}_{\rm b}}{\partial t}+H\vec{v}_{\rm b}+\frac{(\vec{v}_{\rm b}\cdot\nabla)\vec{v}_{\rm b}}{a}+\frac{c_{\rm b}^2}{a}\nabla \delta_{\rm b}&=\frac{(\nabla\times\vec{B})\times\vec{B}}{4\pi a^5\rho_{\rm b}}-\frac{\nabla\phi}{a},\label{eq:thetab}
	\end{align}
	where $H=\frac{d\ln a}{dt}$ is the Hubble rate, $\rho_{\rm b}$ is the energy density of baryons, $c_{\rm b}$ is the baryon sound speed, $\delta_{\rm b}$ and $\phi$ is the metric potential which satisfies the Newtonian Poisson's equation,
	\begin{align}\label{eq:poisson}
		\nabla^2\phi&=\frac{1}{2M_{\rm pl}^2}a^2(\rho_{\rm b}\delta_{\rm b}+\rho_{\rm DM}\delta_{\rm DM}).
	\end{align}
	Here $M_{\rm pl}=2.435\times 10^{18}$ GeV is the reduced Planck mass, $\delta=(\rho(x)-\bar{\rho})/\bar{\rho}$ is the fluid density perturbation, and the subscript DM refers to dark matter.
	
	On comoving length scales, $l$, where baryon velocity is small, $v_{\rm b}/l\ll aH$, the RHS in the induction equation (Eq.~\eqref{eq:induction}) can be ignored, and the comoving magnetic field on those scales remains unchanged. Moreover, on these scales, even the convective term ($(\vec{v}_{\rm b}\cdot\nabla)\vec{v}_{\rm b}$) in the baryon Euler equation (Eq.~\eqref{eq:thetab}) can be neglected. Thus, on large scales, baryons perturbatively follow the Lorentz force from the PMFs while providing negligible feedback to PMFs.\footnote{When applying cosmological perturbation theory with PMFs, one treats $\vec{B}$ itself as a perturbed variable but it's order is lower than the first order matter perturbations, $\delta$ and $v$. For instance, from Eq.~\eqref{eq:thetab}, one can see that $v_{\rm b}$ is of order $B^2$. Moreover, unlike other cosmological perturbations, PMFs also affect the background expansion of the universe via its homogeneous energy density, $\rho_{\rm b}=\langle \vec{B}_{\rm phys}^2\rangle/[8\pi]$. However, as $\rho_{\rm b}$ is at least $10^{-6}$ factors suppressed compared to the photon energy density for PMF strengths of our interest ($B\lesssim$ nG), one can safely neglect $\rho_{\rm b}$ in the Hubble rate.} 
	
	On small scales where $v_{\rm b}/l> aH$, baryon motion can no longer be treated as perturbative, and the feedback of baryons on PMFs coupled with convective terms in the baryon Euler equation leads to magnetohydrodynamic turbulence \cite{Trivedi:2018ejz, Banerjee:2004df, Jedamzik:2018itu}. Turbulence leads to the suppression of both $\delta_{\rm b}$ and $B$ on these small scales.\footnote{Note that while PMFs suppress baryon perturbations below $\lambda_D$, they can significantly enhance dark matter perturbations on these small scales purely through gravitational influence \cite{Ralegankar:2023pyx}.}
	
	One can analytically estimate this damping scale, $\lambda_{\rm D}$, below which both $\delta_{\rm b}$ and $B$ are damped. Considering baryon flow is driven by the Lorentz force (see Eq.~\eqref{eq:thetab}), we have $v_{\rm b}\sim \frac{1}{aH\lambda_{\rm D}}\frac{\vec{B}^2_{\rm phys}}{4\pi\rho_{\rm b}}$. Then by setting $v_{\rm b}/\lambda_{\rm D}\sim aH$ we obtain
	\begin{align}\label{eq:lambdaD_va}
		\lambda_{\rm D}\sim \frac{v_{\rm A}}{aH}.
	\end{align}
	Here $v_{\rm A}$ is the Alfven velocity of the plasma,
	\begin{align}
		v_{\rm A}^2\equiv\frac{\langle \vec{\rm B}^2_{\rm phys}\rangle }{4\pi\rho_{\rm b}},
	\end{align}
	and $\langle..\rangle$ denotes ensemble averaging over all stochastic realisations. Thus, MHD turbulence occurs on comoving length scales traversed by a particle moving with Alfven speed in Hubble time.
	
	In this study, we shall primarily focus on length scales larger than $\lambda_{\rm D}$ where the effects from turbulence can be ignored. In a matter-dominated universe, $\lambda_{\rm D}$ is constant if $B\equiv \sqrt{\langle \vec{B}^2\rangle}= a^2\sqrt{\langle \vec{B}^2_{\rm phys}\rangle}$ is constant. Since $\lambda_{\rm D}$ does not grow and cause damping of larger scale magnetic fields, the assumption of constant $\langle \vec{B}^2\rangle$ is self-consistent. Using present-day values of $\rho_{\rm m0}\approx 1.15\times 10^{-47} {\rm GeV}^4$ in the Hubble rate and $\rho_{\rm b0}\approx 1.64\times10^{-48} {\rm GeV}^4$ in the Alfven velocity \cite{Planck:2018jri}, we obtain
	\begin{align}\label{eq:lambda_sim}
		\lambda_{\rm D}\sim 0.1{\rm Mpc} \left(\frac{B}{\rm nG}\right).
	\end{align}
	
	\subsection{Linear growth of matter density perturbations sourced by PMFs}\label{sec:grow}
	In this study, we shall focus on PMFs with $B>0.1$ nG to produce a significant impact on potentially observable halos with $M_{\rm halo}>10^8 {\rm M}_{\odot}$. Correspondingly, the scales of our interest where baryon perturbations can be treated linearly, $l>\lambda_{\rm D}>$10 kpc, would have negligible baryon thermal pressure. Thus the evolution equation for $\vec{v}_{\rm b}$ is simply given by
	\begin{align}
		\frac{\partial \vec{v}_{\rm b}}{\partial t}+H\vec{v}_{\rm b}&=\frac{1}{a^2}\frac{(\vec{\nabla}\times\vec{B})\times\vec{B}}{4\pi a^3\rho_{\rm b}}-\frac{\vec{\nabla}\phi}{a}. \label{eq:vb}
	\end{align}

	The compressible component of baryon flow determines the evolution of $\delta_{\rm b}$,
	\begin{align}\label{eq:contb}
		\frac{\partial \delta_{\rm b}}{\partial t}+\frac{\vec{\nabla}\cdot \vec{v}_{\rm b}}{a}&=0.
	\end{align}
	To directly see how PMFs cause an increase in $\delta_{\rm b}$, we need to replace $\vec{v}_{\rm b}$ using Eq.~\eqref{eq:vb}. We do so by first taking the time derivative of Eq.~\eqref{eq:contb}, then replacing $\frac{\partial \vec{v}_{\rm b}}{\partial t}$ using Eq.~\eqref{eq:vb} and $\nabla\cdot\vec{v}_{\rm b}$ using Eq.~\eqref{eq:contb}. Then writing the equation in terms of scale factor instead of time, we obtain
	\begin{align}\label{eq:delta_b}
		a^2\frac{\partial^2 \delta_{\rm b}}{\partial a^2}+a\frac{3}{2}\frac{\partial \delta_{\rm b}}{\partial a}=-\frac{S_0}{a^3H^2}+\frac{\nabla^2\phi}{(aH)^2},
	\end{align}
	where $S_0$ is the source term from magnetic fields,
	\begin{align}\label{eq:S0}
		S_0=\frac{\nabla\cdot[(\nabla\times\vec{B})\times\vec{B}]}{4\pi a^3\rho_{\rm b}}.
	\end{align}
	One can see that for a matter-dominated universe, where $a^3H^2$ is constant, the magnetic fields can enhance baryon density perturbations if $\vec{B}$ remains constant.
	
	Note that the evolution of dark matter and baryon perturbations need to be solved together as they gravitationally influence each other. The evolution equation of $\delta_{\rm DM}$ is the same as Eq.~\eqref{eq:delta_b} except the $S_0$ term is absent. Furthermore, by replacing $\phi$ using Poisson's equation (Eq.~\eqref{eq:poisson}) and neglecting dark energy density in the Hubble rate, we obtain
	\begin{align}\label{eq:deltab_a}
		a^2\frac{\partial^2 \delta_{\rm b}}{\partial a^2}+a\frac{3}{2}\frac{\partial \delta_{\rm b}}{\partial a}-\frac{3}{2}\frac{\Omega_{\rm b}}{\Omega_{\rm m}(1+a_{\rm eq}/a)}\delta_{\rm b}=-\frac{S_0}{a^3H^2}+\frac{3}{2}\frac{\Omega_{\rm DM}}{\Omega_{\rm m}(1+a_{\rm eq}/a)}\delta_{\rm DM}\\
		a^2\frac{\partial^2 \delta_{\rm DM}}{\partial a^2}+a\frac{3}{2}\frac{\partial \delta_{\rm DM}}{\partial a}-\frac{3}{2}\frac{\Omega_{\rm DM}}{\Omega_{\rm m}(1+a_{\rm eq}/a)}\delta_{\rm DM}=\frac{3}{2}\frac{\Omega_{\rm b}}{\Omega_{\rm m}(1+a_{\rm eq}/a)}\delta_{\rm b}.\label{eq:deltadm_a}
	\end{align}
	Here $\Omega_{\rm b}=0.048$ is the baryon energy density fraction today, $\Omega_{\rm DM}=0.261$ is the dark matter fraction, $\Omega_{\rm m}=\Omega_{\rm b}+\Omega_{\rm DM}$, and $a_{\rm eq}=2.94\times 10^{-4}$ is the matter-radiation equality \cite{Planck:2018jri}.
	
	The above system of equations can be solved once the initial conditions and the value of $S_0$ is specified. As the above equations are ordinary differential equations, their solution can simply be written as a linear combination of the homogeneous solution provided by the initial condition and the inhomogeneous solution sourced by $S_0$, $\delta=\delta^{\rm \Lambda CDM}+\delta^{\rm PMF}$. The solution of $\delta$ sourced just by initial conditions, $\delta^{\rm \Lambda CDM}$, will be the same as in standard cosmology, and hence we will not focus on those solutions in this section.
	
	\begin{figure}
		\begin{subfigure}{0.5\textwidth}
			\includegraphics[width=1.00\textwidth]{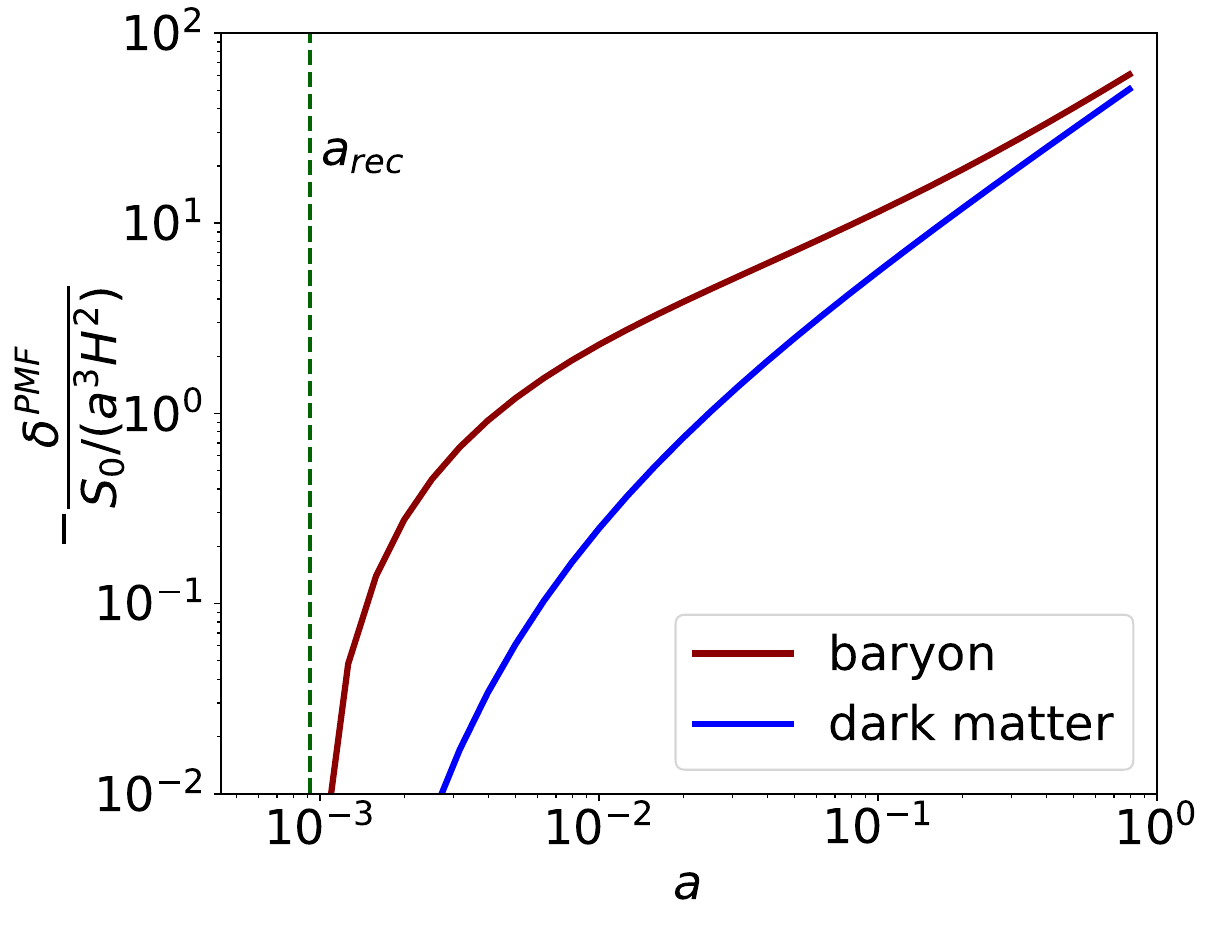}
		\end{subfigure}
		\begin{subfigure}{0.5\textwidth}
			\includegraphics[width=1.00\textwidth]{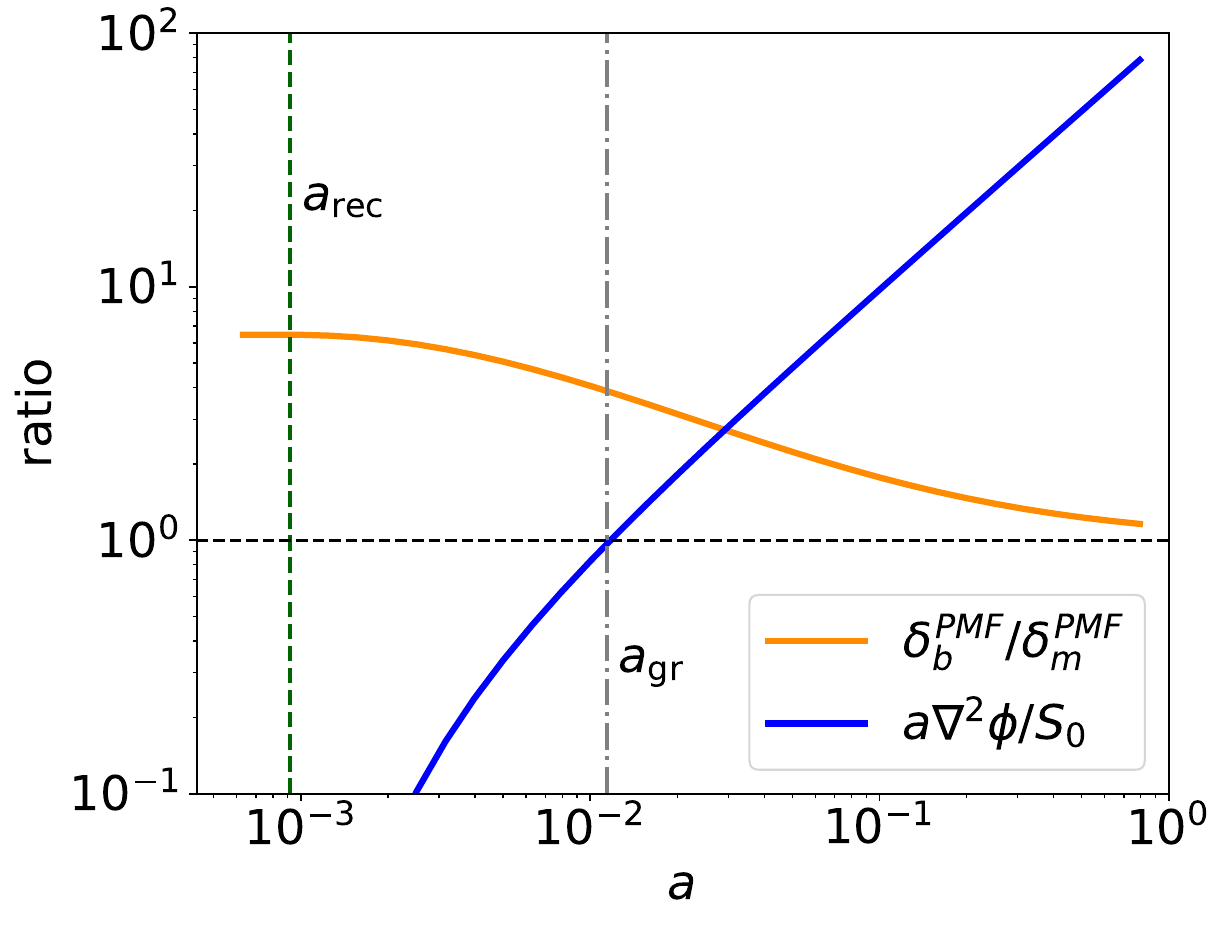}
		\end{subfigure}
		\caption{\textbf{Left}: Evolution of baryon (dark red) and dark matter (blue) density perturbations normalised with $S_0/[a^3H^2]$, which parameterises the Lorentz force. \textbf{Right}: Blue line shows the ratio of gravitational force with the Lorentz force acting on baryons. Orange line represents the ratio of baryon fraction in an overdensity compared to the cosmic average $\Omega_{\rm b}/\Omega_{\rm m}$. Both the left and right panels are scale-independent. This is because all the scale dependence is sourced by the $S_0$ term and since we are plotting ratios the scale dependence is removed.}\label{fig:xi}
	\end{figure}
	
	The value of $\delta^{\rm PMF}$ prior to recombination is suppressed due to large photon drag experienced by baryons. Thus, we take the initial condition at $a=a_{\rm rec}$ as $\delta^{\rm PMF}_{\rm b}=\delta^{\rm PMF}_{\rm DM}=0$ and $v^{\rm PMF}_{\rm b}=v^{\rm PMF}_{\rm DM}=0$. 

 After recombination, the evolution of $\delta^{\rm PMF}$ is determined by the $S_0$ term. Specifically, we can write the post-recombination evolution of density perturbations as
	\begin{align}\label{eq:xi_def}
		\delta_{\rm b}^{\rm PMF}=-\xi_{\rm b}(a)\frac{S_0}{a^3H^2} && \delta_{\rm DM}^{\rm PMF}=-\xi_{\rm DM}(a)\frac{S_0}{a^3H^2}.
	\end{align}
	Here $\xi_{\rm b}$ and $\xi_{\rm DM}$ are dimensionless factors that contain the time-evolution of density perturbations and are independent of the strength of PMFs as well as the scales under consideration, as long as the scale is larger than $\lambda_{\rm D}$. 
 
    In the left panel of Fig.~\ref{fig:xi} we show the evolution of $\xi$ (see also \cite{Shibusawa:2014fva}). One can see that $S_0$ causes growth in $\delta_{\rm b}$, which in turn gravitationally sources growth in $\delta_{\rm DM}$. For the most part of the evolution, $\delta_{\rm DM}$ is lagging behind $\delta_{\rm b}$.
	
	If these matter perturbations collapse into a halo, then they will naturally have a larger baryon fraction. One can estimate the baryon fraction in the halos at any redshift simply as,
	\begin{align}\label{eq:fb}
		f_{\rm b}=\frac{\delta\rho_{\rm b}}{\delta\rho_{\rm DM}+\delta\rho_{\rm b}}=\bar{f}_{\rm b}\frac{\delta_{\rm b}}{\bar{f}_{\rm DM}\delta_{\rm DM}+\bar{f}_{\rm b}\delta_{\rm b}}\equiv \bar{f}_{\rm b}\frac{\delta_{\rm b}}{\delta_{\rm m}},
	\end{align}
	where $\bar{f}_{\rm b}=\Omega_{\rm b}/\Omega_{\rm m}\approx0.154$ and $\bar{f}_{\rm DM}=\Omega_{\rm DM}/\Omega_{\rm m}\approx0.845$ are the cosmic averaged values of baryon and DM fraction \cite{Planck:2018jri}, and $\delta_{\rm m}$ is the matter density perturbation.
	In the right panel of Fig.~\ref{fig:xi}, we show the evolution of $f_{\rm b}/\bar{f}_{\rm b}$ with scale factor. One can see that for halos collapsing at $z\sim 10$, one can expect a baryon fraction in halos about two times larger than the cosmic average. Note that this enhancement is independent of the halo mass. Later when we consider the influence of PMFs along with inflationary initial conditions, we see mass dependence in the enhancement of $f_{\rm b}$. 
	
	There are two forces inducing growth in $\delta_{\rm b}$: gravity and the Lorentz force. In the left panel of Fig.~\ref{fig:xi}, we show the evolution of the gravitational force $\nabla^2\phi$ compared to the source term $S_0$ from PMFs. One can see that gravity quickly dominates over the Lorentz force by $10a_{\rm rec}$.
	
	To understand how $\delta_{\rm b}$ evolves only under Lorentz force, let us focus on the initial evolution when $\delta_{\rm b}$ and $\delta_{\rm DM}$ are much smaller than $S_0/[a^3H^2]$. In this limit, the terms with $\Omega$ in Eq.~\eqref{eq:deltab_a} can be ignored and one obtains a logarithmic growth for $\delta_{\rm b}$,
	\begin{align}
		\delta_{\rm b}\approx 2\frac{S_0}{a^3H^2}\log(a/a_{\rm rec}). 
	\end{align}
	
	This logarithmic growth continues until self-gravity from baryons becomes  important, i.e. when $\frac{3}{2}\frac{\Omega_{\rm b}}{\Omega_{\rm m}}\delta_{\rm b}\sim \frac{S_0}{a^3H^2}$. Substituting the above logarithmic growth of $\delta_{\rm b}$, the scale factor when gravity takes over, $a_{\rm gr}$, is then simply given by
	\begin{align}
		a_{\rm gr}\sim a_{\rm rec}\exp\left(\frac{\Omega_{\rm m}}{3\Omega_{\rm b}}\right)\approx 10a_{\rm rec}.
	\end{align}
	
	Thus, by redshift $z=100$, the value of $\delta_{\rm b}$ becomes large enough for gravity to take over.  It is important to note that the time when gravity overcomes the Lorentz force is the same for all length scales (larger than $\lambda_{\rm D}$) and is independent of the strength of PMFs.
	
	\subsubsection{Remark on magnetic Jeans scale} \label{subsec:Jeans}
	The above description of gravity overcoming magnetic fields after $a>10a_{\rm rec}$ naively seems to be at odds with the main narrative in literature where the dominance of gravity is expressed in terms of the magnetic Jeans scale. Specifically, the magnetic Jeans scale, $\lambda_{\rm J}$, is evaluated by finding the scale where the magnetic pressure balances the gravitational collapse. Taking analogy with the thermal Jeans scale and replacing the sound speed with Alfvén speed, Ref.~\cite{Kim:1994zh} finds the comoving magnetic Jeans scale to be
	\begin{align}
		\lambda_{\rm J}=\frac{v_{\rm A}}{a}\sqrt{\frac{\pi}{G\rho_{\rm m}}}=2\pi\sqrt{\frac{2}{3}}\frac{v_{\rm A}}{aH}.
	\end{align}
	In the last line, we used the fact that in a matter-dominated universe, $H^2=8\pi G\rho_{\rm m}/3$.
	
	The reasoning behind the magnetic Jeans scale is somewhat incomplete because it suggests that magnetic fields only act to oppose the growth of density perturbations. However, as we saw in the previous section, PMFs for the large part support the growth of density perturbations via the $S_0$ term. It is only when the baryon perturbations become large enough to backreact onto the magnetic fields, that magnetic fields suppress density perturbations by inducing turbulence.
	% However, such a narrative is somewhat flawed because PMFs are not opposing the growth of density perturbations but are, in fact, supporting it. The narrative of the magnetic jeans scale is more appropriate for scenarios with homogenous magnetic fields, where the $S_0$ term sourcing density perturbation vanishes.
	
	Interestingly, the damping scale $\lambda_{\rm D}$ (Eq.~\eqref{eq:lambdaD_va}), which marks the scale where baryons back-react onto PMFs, is of the same form as $\lambda_{\rm J}$ above. This similarity is surprising because gravity had no role in the derivation of $\lambda_{\rm D}$. However, gravity does play an indirect role in the overall narrative behind $\lambda_{\rm D}$ once we consider the evolution of baryon perturbations on scales larger than $\lambda_{\rm D}$. On these scales, the baryon perturbations sourced by $S_0$ are small and only grow logarithmically initially. It is gravity that ultimately escorts the perturbations to non-linear values. As the strength of gravity is much larger than the Lorentz force from PMFs by the time the perturbations become non-linear, the back-reaction from magnetic fields cannot inhibit the collapse of structures. Whereas perturbations on scales smaller than $\lambda_D$ are suppressed by MHD turbulence before gravity overcomes the Lorentz force. Thus, $\lambda_{\rm D}$ effectively plays the same role as the magnetic Jeans scale. 
	
	\subsection{Power spectrum with PMFs}\label{sec:power}
	To obtain the power spectrum of baryons and dark matter sourced by PMFs, one would first need to specify the spectrum for PMFs. In this study, we focus on non-helical PMFs for simplicity, where
	\begin{align}\label{eq:def_M}
		\langle B_i(k) B^*_j(k')\rangle=(2\pi)^3\delta^3\!(k-k')\left(\delta_{ij}-\frac{k_ik_j}{k^2}\right)\frac{P_{\rm B}(k)}{2}.
	\end{align}
	Above $B_i(k)$ is the Fourier transform of $B_i(x)$, where we use the following convention for Fourier transforms: $A(k)=\int d^3x A(x) e^{ikx}$.
	
	Furthermore, we shall consider $P_{\rm B}$ to be of the form
	\begin{align}\label{eq:PB_powerlaw}
		P_{\rm B}(k)=Ak^{n_{\rm B}}e^{-k^2\lambda_{\rm D}^2}.
	\end{align}
	Above, the exponential term is added to model the damping from the back-reaction of baryons below the scale $\lambda_{\rm D}$.\footnote{In reality, MHD turbulence would cause a power-law damping of PMFs below the scale $\lambda_{\rm D}$ \cite{Banerjee:2004df, Trivedi:2018ejz}. In this study, we focus on scenarios where the evolution of large-scale modes is insensitive to the behaviour on scales smaller than $\lambda_{\rm D}$ and hence the difference in modelling of damping is inconsequential.}
	The amplitude $A$ is determined by specifying the strength of PMFs averaged over $\lambda_{\rm Mpc}=1$ Mpc,
	\begin{align}\label{eq:B1mpc}
		B^2_{\rm 1 Mpc}\equiv\int \frac{d^3k}{(2\pi)^3}P_{\rm B}(k)e^{-k^2\lambda_{\rm Mpc}^2}= \frac{A\lambda_{\rm Mpc}^{-(3+n_{\rm B})}}{4\pi^2}\Gamma([n_{\rm B}+3]/2),
	\end{align}
	where to obtain the second relation we assumed $\lambda_{\rm Mpc}\gg \lambda_{\rm D}$ and $n_{\rm B}>-3$ so that the integral is not sensitive to large and small scale cutoffs. Here $\Gamma$ is the Gamma function. One can also calculate the total strength of PMFs, $B^2$, by removing $e^{-k^2\lambda_{\rm Mpc}^2}$ from the above integral. The total PMF strength can then be related to $B_{\rm 1Mpc}$ using
	\begin{align}\label{eq:B_B1mpc}
		B^2=B_{\rm 1Mpc}^2\left(\frac{\lambda_{\rm Mpc}}{\lambda_{\rm D}}\right)^{n_{\rm B}+3}.
	\end{align}
	
	Having parameterised the spectrum of PMFs, we now come back to the evaluation of the baryon power spectrum induced by PMFs, $P_{\rm b}^{\rm PMF}$. From Eq.~\eqref{eq:xi_def}, one can see that  $P_{\rm b}^{\rm PMF}$ can be directly determined from the power spectrum of $S_0$. Taking the Fourier transform of $S_0$ given in Eq.~\eqref{eq:S0} and neglecting non-gaussianities in PMFs while taking the ensemble average $\langle \delta_{\rm b}(k) \delta_{\rm b}^*(k')\rangle$, we obtain (see also \cite{Kim:1994zh, Adi:2023doe})
	\begin{multline}\label{eq:Pb_int}
		P^{\rm PMF}_{\rm b}(k) =\xi_{\rm b}^2(a)\frac{k^4}{8(4\pi a^3\rho_{\rm b}[a^3H^2])^2}\int \frac{d^3q  }{(2\pi)^3}\frac{P_{B}(q)P_{\rm B}(k-q)}{(k-q)^2}\bigg[k^2+2q^2+4\frac{(q\cdot k)^4}{k^4q^2}\\-4\frac{(q\cdot k)^2}{k^2}-4\frac{(q\cdot k)^3}{k^2q^2}+\frac{(q\cdot k)^2}{q^2}\bigg].
	\end{multline}
	For $P_{\rm B}(q)\propto q^{\rm n_{\rm B}}$ with $n_{\rm B}>-3/2$, the above integral diverges at large $q$ values and hence is sensitive to the small-scale cutoff, $\lambda_{\rm D}$. Specifically, $P_{\rm b}^{\rm PMF}\propto k^4$ with the proportionality constant being sensitive to both $n_{\rm B}$ and $\lambda_{\rm D}$. As the exact spectrum of PMFs near $k\sim \lambda_{\rm D}^{-1}$ has large theoretical uncertainties, we do not focus on scenarios with $n_{\rm B}
	>-3/2$ in this study.
	
	For $-3<n_{\rm B}<-1.5$, the integral in Eq.~\eqref{eq:Pb_int} is largely sensitive to $q\sim k$ and hence is decoupled from both the small and large scale cut-offs. Considering $P_{\rm B}=Ak^{\rm n_{\rm B}}$ in Eq.~\eqref{eq:Pb_int}, replacing $A$ using Eq.~\eqref{eq:B1mpc}, integrating over $q$, and using present-day values of $\rho_{\rm m0}\approx 1.15\times 10^{-47} {\rm GeV}^4$ in the Hubble rate and $\rho_{\rm b0}\approx 1.64\times10^{-48} {\rm GeV}^4$ we obtain
	\begin{align}\label{eq:Delta_b}
		\Delta_{\rm b}^{\rm PMF}(k)\equiv\frac{k^3P_{\rm b}^{\rm PMF}(k)}{2\pi^2}=10^{-4}\xi_{\rm b}^2(a) \left(\frac{k}{\rm Mpc^{-1}}\right)^{2n_{\rm B}+10} \left(\frac{B_{\rm 1 Mpc}}{\rm nG}\right)^4G_{\rm n_{\rm B}}e^{-2k^2\lambda_{\rm D}^2},
	\end{align}
	where $G_{\rm n_B}$ is a dimensionless number determined by
	\begin{align}
		G_{\rm n_B}=\int_0^{\infty} dx \int _{-1}^1\frac{dy}{2}x^{n_{\rm B}+2}(1+x^2-2xy)^{n_{\rm B}/2-1}\frac{\left[1+2x^2+4y^4x^2-4y^2x^2-4y^3x+y^2\right]}{\Gamma^2([n_{\rm B}+3]/2)}.
	\end{align}
	In Eq.~\eqref{eq:Delta_b} we added the exponential factor of $e^{-2k^2\lambda_{\rm D}^2}$ by hand to model the damping of baryon density perturbations near $\lambda_{\rm D}$. The power spectrum of dark matter is of the same form with $\xi_{\rm b}$ replaced by $\xi_{\rm DM}$.

    The total baryon (or dark matter) power spectrum can be found by using the fact that the PMF-induced perturbations can be super-imposed on the $\Lambda$CDM perturbations (see also \cite{Kunze:2022mlr}),
    \begin{align}
        \delta=\delta^{\rm PMF}+\delta^{\rm \Lambda CDM}.
    \end{align}
	The above is valid as long as perturbations are small and in the linear limit. 
	
	Furthermore, one naturally expects $\delta^{\rm PMF}$ to be uncorrelated with $\delta^{\rm \Lambda CDM}$. This is because in typical inflationary magnetogenesis mechanisms, the PMFs are sourced by the vacuum fluctuations of the gauge field itself and not the inflaton \cite{Turner:1987bw,Ratra:1991bn}. Consequently, PMFs are uncorrelated with inflationary curvature perturbations. The PMFs generated from first order phase transitions are typically generated from turbulent processes and are also expected to be uncorrelated with inflationary curvature perturbations. Thus, the total power spectrum can simply be written as,
    \begin{align}
        P(k)=P^{\rm PMF}(k)+P^{\rm \Lambda CDM}(k).
    \end{align}
    The above is always true for dark matter but only for $k<\lambda_D^{-1}$ for baryons. For $k>\lambda_D^{-1}$, the baryon perturbations enter the non-linear regime and the total baryon power spectrum is expected to be suppressed due to turbulence.
    
	\begin{figure}
		\begin{subfigure}{0.5\textwidth}
			\includegraphics[width=1.00\textwidth]{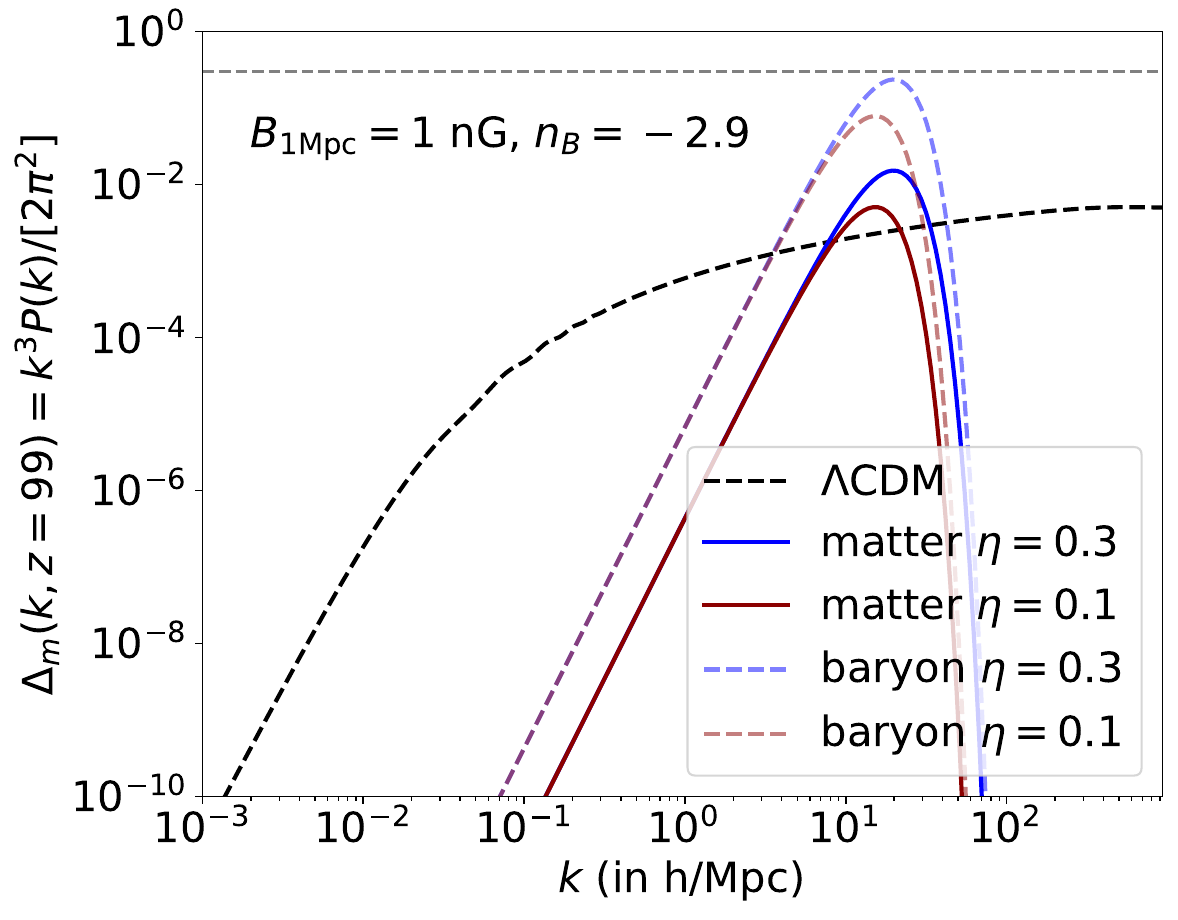}
		\end{subfigure}
		\begin{subfigure}{0.5\textwidth}
			\includegraphics[width=1.00\textwidth]{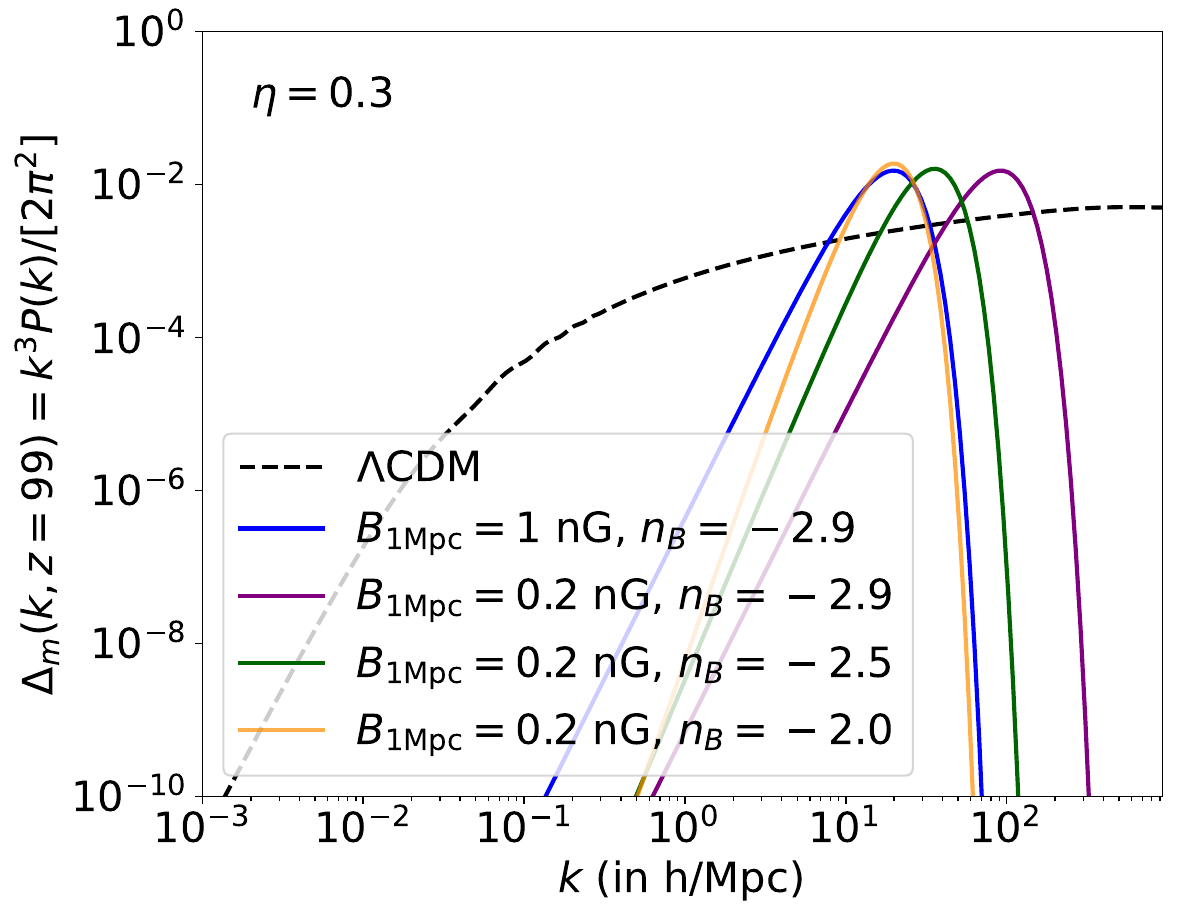}
		\end{subfigure}
		\caption{Comparison of dimensionless matter power spectra from PMFs (coloured solid lines) with that from $\Lambda$CDM cosmology (black dashed line). The coloured dashed lines in the left panel show baryon power spectra. The cutoff scale is determined by when the baryon power spectrum becomes equal to $\eta$. Increasing $B_{\rm 1Mpc}$ or $n_{\rm B}$ simply shifts the cutoff to a larger scale but keeps the maximum value of the power spectrum fixed.}\label{fig:Pk_an}
	\end{figure}
	
	For $n_{\rm B}>-3$, which is required to keep PMFs insensitive to large-scale cutoff, we find the dimensionless baryon power spectrum sourced by PMFs is blue-tilted. In contrast, the dimensionless matter-power spectrum, $\Delta_{\rm m}$, in $\Lambda$CDM cosmology only grows logarithmically on small scales. Thus, PMFs can naturally enhance the small-scale matter power spectrum while keeping the large scales unchanged. In Fig.~\ref{fig:Pk_an}, we compare $\Delta_{\rm m}$ obtained from PMFs with $\Delta_{\rm m}$ from $\Lambda$CDM cosmology. 
	
	Note that while the power-law amplitude of $\Delta_{\rm b}$ in Eq.~\eqref{eq:Delta_b} is sensitive to the strength of PMFs, its peak value is insensitive to $B$. One can see this algebraically by setting $k=\lambda_{\rm D}^{-1}$ in Eq.~\eqref{eq:Delta_b} and replacing $B_{\rm 1Mpc}$ in terms of $B$ using Eq.~\eqref{eq:B_B1mpc}. The resulting equation contains a factor of $(B/\lambda_{\rm D})^4$, which is a constant as $\lambda_{\rm D}$ is proportional to $B$, see Eq.~\eqref{eq:lambda_sim}. Parameterizing the proportionality constant with $\kappa_{\rm n_B}$, such that
	\begin{align} \label{eq:lamkap}
		\lambda_{\rm D}=0.1\kappa_{\rm n_B} {\rm Mpc} \left(\frac{B}{\rm nG}\right),
	\end{align}
	we find
	\begin{align}
		\Delta_{\rm b}^{\rm PMF}(k=\lambda_{\rm D}^{-1})=\xi_{\rm b}^2(a)\kappa_{\rm n_B}^{-4}G_{\rm n_B}e^{-2}.
	\end{align}
	For $\kappa_{\rm n_B}\sim \mathcal{O}(1)$ and at $a=0.01$ where $\xi_b\approx 2.5$, one finds that $\Delta_{\rm b}(\lambda_{\rm D})\sim \mathcal{O}(1)$. The order one value of $\Delta_{\rm b}$ is not a coincidence and is simply a result of the fact that $\lambda_{\rm D}$ is determined by the scales where PMFs induce non-linear fluctuations in baryon perturbations.
	
	It is important to note that the maximum value of $\Delta_{\rm b}$ is sensitive to $\kappa_{\rm n_B}^{-4}$. Thus, even a factor 3 change in $\kappa_{\rm n_B}$ can introduce 2 orders of magnitude change in the maximum value of $\Delta_{\rm b}$. This large theoretical uncertainty in the maximum value of $\Delta_{\rm b}$ has not been taken into account in previous literature. Hence, the earlier constraints on PMFs derived from PMFs' impact on the matter power spectrum could change substantially once the correct value of $\kappa_{\rm n_B}$ is determined.
	
	Determining the exact value of $\kappa_{\rm n_B}$ is beyond the scope of this work as it requires detailed MHD simulations that include gravity.
	In this study, we set the value of $\kappa_{\rm n_B}$ by requiring that the baryon power at the cut-off scale is equal to some order one number parameterized by $\eta$ when gravity overcomes the Lorentz force,
	\begin{align}\label{eq:eta}
		\Delta_{\rm b}^{\rm PMF}(k=\lambda_{\rm D}^{-1}, a=0.01)=\eta.
	\end{align}
	Using $\xi_{\rm b}(a=0.01)\approx 2.5$, we obtain
	\begin{align}\label{eq:kappa_val}
		\kappa_{\rm n_B}= \left(\frac{G_{\rm n_B}}{1.14\eta}\right)^{1/4}.
	\end{align}
	In the left panel of Fig.~\ref{fig:Pk_an}, we show both the baryon and matter power spectrum for two different values of $\eta$. The matter power spectrum is smaller than the baryon power spectrum because $\delta_{\rm DM}\ll \delta_{\rm b}$ when perturbations are sourced by PMFs.
	
	In the right panel, we show the matter power spectrum for different values of $B_{\rm 1Mpc}$ and $n_{\rm B}$ but keep $\eta$ fixed to 0.3. One can see that the peak value of the matter power spectrum remains unchanged for different PMFs parameters but the scale at which PMFs enhance the power spectrum changes. In contrast, previous literature \cite{Subramanian:1997gi, Sethi:2004pe, Shaw:2010ea, Adi:2023doe} found the peak of the power spectrum to change with $n_{\rm B}$ because they neglected the $n_{\rm B}$ dependence in $\lambda_{\rm D}$. For instance, when $\eta=0.3$ in Eq.~\eqref{eq:kappa_val}, we obtain: $\kappa_{-2.9}\approx0.67$, $\kappa_{-2.5}\approx1.07$, $\kappa_{-2}\approx1.36$, and $\kappa_{-1.6}\approx 2.03$. In contrast, the naive $\lambda_{\rm D}$ employed by earlier studies assumed $n_{\rm B}$ independent cutoff scale with $\kappa=0.7$ \cite{Subramanian:1997gi} or $\kappa=0.35$ \cite{Shaw:2010ea}.
	
	To take into account the uncertainty in the exact estimation of $\lambda_{\rm D}$, in this study, we show the results for two values of $\eta$: $\eta=0.3$ and $\eta=0.1$. We avoid using $\eta$ values larger than 0.3 because, in those cases, the assumption of linear perturbations during the generation of initial conditions fails, leading to unwanted numerical artefacts. For more details on the choice of $\eta$, see appendix~\ref{sec:eta_fix}.
	
	\section{Structure formation with PMFs in the initial conditions}\label{sec:sim}
	In this study, we are primarily interested in obtaining the properties of dark matter halos at high redshifts ($z\sim 4-10$), while focusing on their masses, baryon fractions, and abundance of stars. To find these quantities and their relations, we resort to cosmological simulations. For this purpose, we use \texttt{P-Gadget-3}, an extended, non-public version of the publicly available \texttt{Gadget-2} code \cite{VS_gadget2}. In the following, we will refer to it as `Gadget'.\footnote{There is also a recent, updated version \texttt{Gadget-4}, that shares many features with our \texttt{P-Gadget-3} code. It is comprehensively presented in \cite{gadget4}, whereby in Chapter 11 one can find a discussion on the development of the Gadget code and its public availability.}
	
	Our implementation of the Gadget code carries out gravitational N-body collisionless dynamics for dark matter (DM) and smoothed particle hydrodynamics (SPH) for baryonic gas, including cooling of the gas and star formation. We emphasize that our simulations do not entail magneto-hydrodynamics (MHD) and, thus, cannot represent the realistic and complex nature of plasma physics. Some recent works used MHD simulations to study the evolution of PMFs through their impact on population of galaxies and galactic \cite{Marinacci:2015ria}, intergalactic \cite{Vazza:2017mbz} and extragalactic \cite{Vazza:2017qge, Pomakov:2022cem} magnetic properties, also in comparison with astrophysical seeding mechanisms during galaxy formation \cite{Garaldi:2021}. However, as discussed in the previous section, on sufficiently large scales, the effects of the Lorentz force become sub-dominant to gravity after $a>10\, a_{\rm rec}$, i.e., for redshift $z<100$. Therefore, in the formation of structures on considered scales, we do not expect a significant difference in the evolution of density perturbations between our simulations and a full MHD picture.
	
	The comoving box sizes ($L_{\rm box}$) of the simulations are set such that the Nyquist frequency $k_{{\rm Nyq}}$ is about 1.5 times larger than the wavenumber corresponding to the damping scale, $k_{\rm D}=1/\lambda_{\rm D}$. Each simulation box is filled with an equal number of gas and DM particles, $N = 512^3$ (with $N= N_{{\rm DM}} = N_{{\rm gas}}$). 
	
	The simulations are performed from initial conditions (ICs), provided at $z_{{\rm in}}=99$, and run down to $z_{{\rm end}}=0$ or $4$ (see Table~\ref{table:simparams}). For the $\Lambda$CDM cosmology, ICs were obtained by computing the $\Lambda$CDM power spectra of baryons and dark matter at $z_{{\rm in}}$, using \texttt{CLASS} code \cite{CLASS}, and inserting those in the \texttt{N-GenIC} software \cite{NGENIC}. The ICs with PMFs (simulations A, D, E, and F) are evaluated by first generating magnetic fields on the lattice for a given PMF power spectrum. Then dark matter and baryon particles are displaced at each grid point according to the sum of Lorentz force, $L_{\rm B}\propto (\vec{\nabla}\times\vec{B})\times \vec{B}$, at that point and the displacement from original $\Lambda$CDM ICs. The resulting ICs are found to have a power spectrum smaller than the analytical estimate. The source of the mismatch is unclear and we leave its detailed investigation to future work. In this study, we amplify the displacement of particles along the Lorentz force direction to match with the analytical estimate.
	%To match the numerical power spectrum with the analytical estimate we further manually modify the ICs. 
	See Appendix~\ref{sec:ngenic} for a more detailed explanation of how the ICs are generated.

 	\begin{table}
		\centering
		\caption{{\scshape Parameters of simulations}. From left to right, columns correspond to: the simulation name, PMF strength at 1 Mpc, spectral index of the PMF power spectrum, parameter setting the peak value of the baryon power spectrum,  periodic box size, the final redshift of the simulation; the last column indicates whether the ICs are generated by the input of the enhanced power spectrum or by the displacement of particles with the Lorentz force. With and without isocurvature refers to whether the PMF-induced perturbations and the $\Lambda$CDM perturbations are uncorrelated or correlated, respectively. Below are listed values of some derived parameters: gravitational softening length, the Nyquist wavenumber, PMF damping wavenumber, and particle masses of DM and gas. At the bottom, we list the values of the total matter energy density, baryon energy density and the dimensionless Hubble parameter (with $H_0 = h \times 100 \frac{\text{km}}{\text{Mpc}\,\text{s}}$). All simulations start from redshift $z_{{\rm in}}=99$, contain $N=512^3$ particles of both types, and have their $\Lambda$CDM counterpart.}
		% $m_{\rm DM} = \rho_c \Omega_{\rm DM} L_{box}^3 / N$, $m_{\rm gas} = \rho_c \Omega_{b} L_{box}^3 / N$, $k_{Nyq} = \pi N^{1/3} / L_{box}$
		\begin{tabular}{c c c c c c c} 
			\hline
			Simulation & $B_{\rm 1Mpc}$ & $n_{B}$ & $\eta$ & $L_{{\rm box}}$  &  $z_{{\rm end}}$ & particle displacement \\ 
			& [nG] & & & [Mpc/$h$] &   & method with PMFs\\
			\hline\hline
			A & 1 & -2.9 & 0.3 & 55 & 0& Lorentz force \\
			B & 1 & -2.9 & 0.3 & 55  & 4& $P(k)$ without isocurvature  \\
			C & 1 & -2.9 & 0.3 & 55  & 4& $P(k)$ with isocurvature\\
			D & 1 & -2.9 & 0.1 & 100  & 0& Lorentz force \\
			E & 0.2 & -2.0 & 0.3 & 55 & 4& Lorentz force  \\
			F & 0.5 & -2.9 & 0.3 & 30  & 4& Lorentz force  \\
			\hline
		\end{tabular} \\
		\begin{tabular}{c}
			Derived parameters
		\end{tabular} \\
		\begin{tabular}{c c c c c c} 
			\hline
			Simulation & $l_{{\rm soft}}$ & $k_{{\rm Nyq}}$ & $k_{\rm D}=1/\lambda_{\rm D}$ & $m_{{}_{\rm DM}}$ & $m_{{\rm gas}}$ \\ 
			& $[{\rm kpc}/h]$  & $[h/{\rm Mpc}]$ & $[h/{\rm Mpc}]$ & $[{\rm M}_{\odot}/h]$ & $[{\rm M}_{\odot}/h]$ \\ [0.5ex] 
			\hline\hline
			$\lbrace {\rm A,B, C} \rbrace$, E  & 4.30 & 29.25  & $\lbrace 19.38 \rbrace $, 16.29 & $8.94\times 10^7$ & $1.66\times 10^7$  \\ 
			D  & 7.81 & 16.08  & 14.93 & $5.37\times 10^8$ & $9.97\times 10^7$ \\ 
			F  & 2.34 & 53.62  & 37.52 & $1.45\times 10^7$ & $2.69\times 10^6$ \\ 
			\hline
		\end{tabular} \\
		\begin{tabular}{c}
			Cosmological parameters (Planck 2015 data \cite{Ade:2015xua})
		\end{tabular} \\
		\begin{tabular}{c c c c c c c c c} 
			\hline
			& & $\Omega_{\rm m} = 1-\Omega_{\Lambda}$ &  & $\Omega_{\rm b} $  &  &  $h$  &  &   \\
			&  & 0.308 &  & 4.82 $\times 10^{-2}$ &  &  0.678  &  &  \\
			\hline
		\end{tabular}
		\label{table:simparams}
	\end{table}

\begin{figure}
    \centering
        \includegraphics[width=0.94\textwidth,trim={0 2.1cm 0 0.1cm},clip]{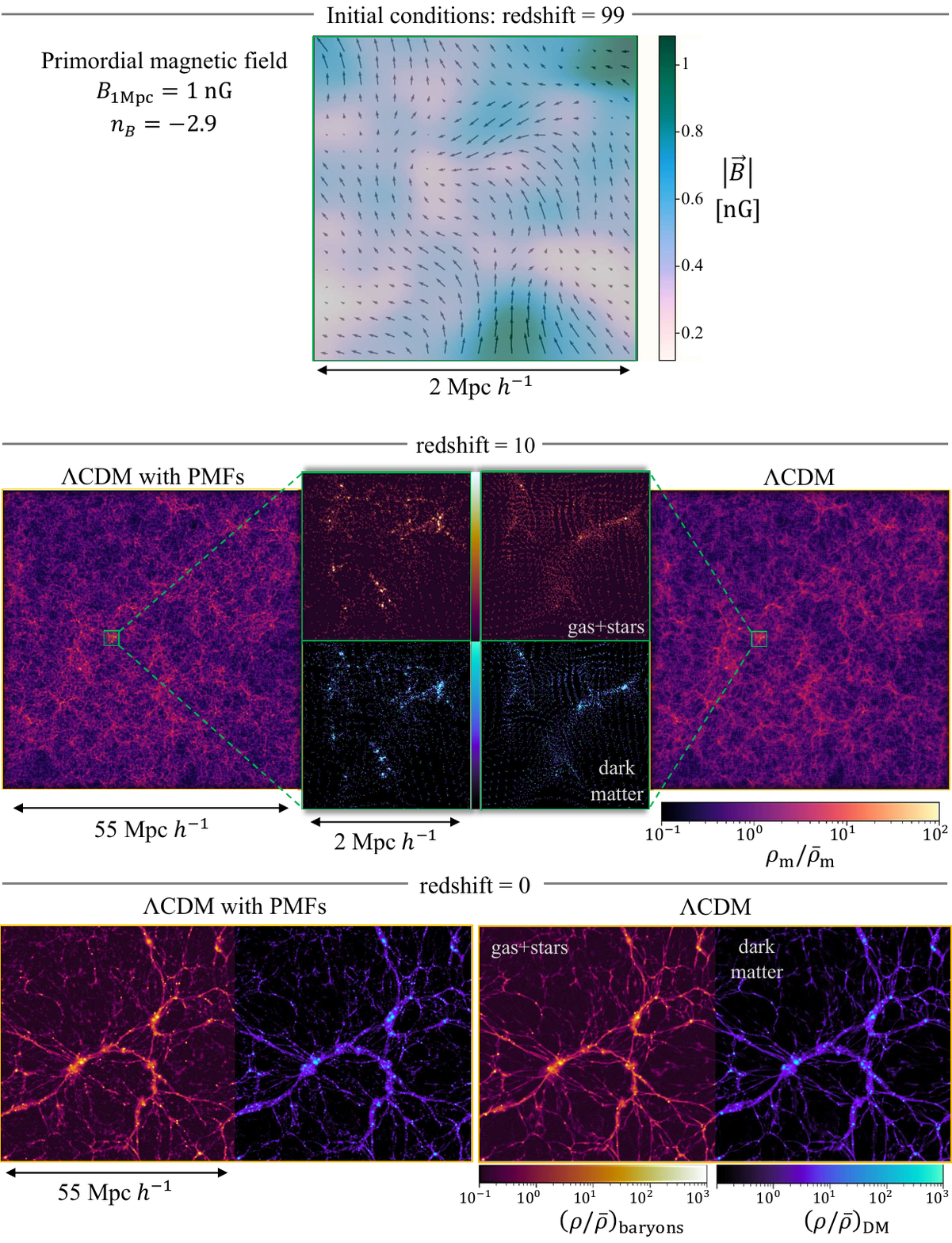}
        
        \caption{A thin slice of scale 0.2 Mpc/$h$ is cut out of the simulation volume and projected onto a 2D plane. {\bf Top:} PMFs in the initial conditions. {\bf Middle:} Zoom-in on a small region at redshift 10 (with same coordinates as in the Top panel), showing small-scale differences of overdensities $\rho/\bar{\rho}$. Color bars in the middle are linearly scaled between 0 and 100. {\bf Bottom:} Overdensity maps of baryons and dark matter at zero redshift in the whole slice.}\label{fig:slice}
    \end{figure}

 In Fig.~\ref{fig:slice} we show an example of a PMF configuration in the ICs of simulation A, and subsequent evolution of baryons and dark matter in the simulation, in comparison with the pure $\Lambda$CDM counterpart. The figure is a 2D projection of a thin slice (depth 0.2 Mpc/$h$) of the whole simulation volume (with $L_{\rm box}=55 $ Mpc/$h$). In the top panel, the portrayed magnetic field vectors are in the middle of the slice and averaged over four mesh layers, while the magnitude of the local magnetic field is coloured in the background. The middle panel demonstrates the small-scale differences between the simulations when PMFs are present versus absent. In particular, one can observe more halos and stars in the PMF simulation at redshift 10. The small-scale differences are also prominent at redshift=0, where more bright spots are visible in the PMF case. In contrast, the large-scale structure is the same in both.
 %Closer examination reveals that, due to PMFs, the large-scale distribution of matter is more scattered and less filamentary compared to the smoother, $\Lambda$CDM-only version. 
 %This detail is more prominent at later redshifts (bottom panel), where more bright spots are visible in the scenario when PMFs are included.
	
\begin{figure}
		\begin{subfigure}{0.5\textwidth}
			\includegraphics[width=1.00\textwidth]{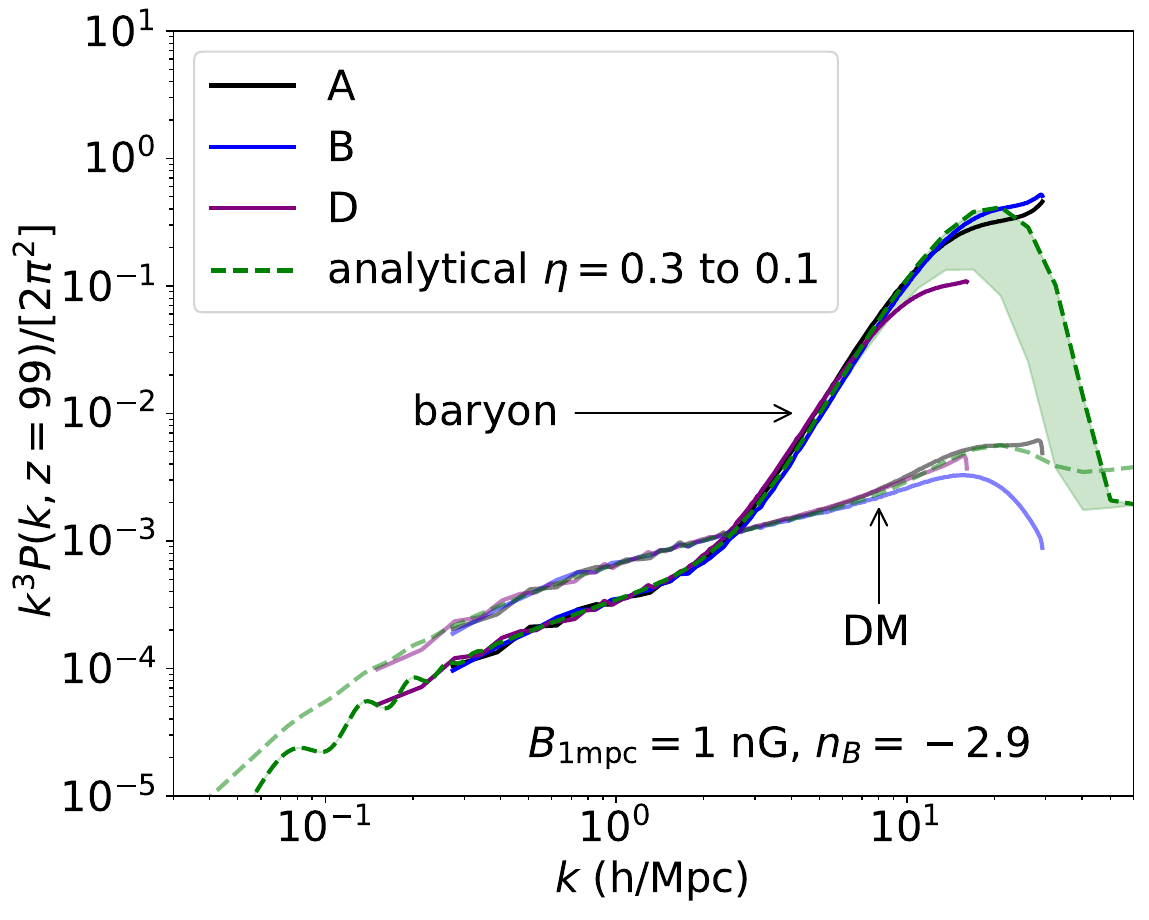}
		\end{subfigure}
		\begin{subfigure}{0.5\textwidth}
			\includegraphics[width=1.00\textwidth]{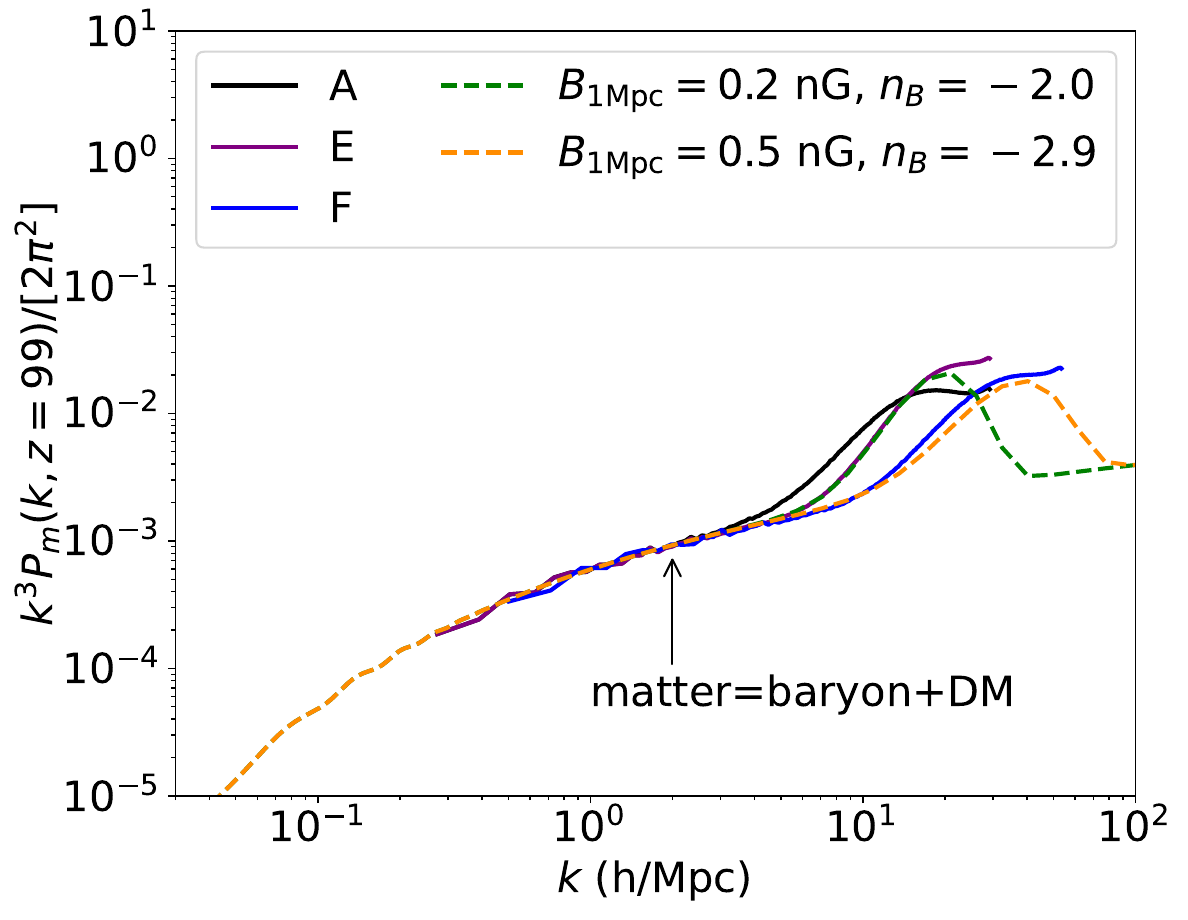}
		\end{subfigure}
		\caption{\textbf{Left:} The solid coloured lines show the dimensionless baryon power spectra of the initial conditions of different simulations. The slightly transparent lines are for dark matter. The power spectra for simulation C are completely identical to that of B and hence are not shown in the left panel. The dashed lines show the analytical power spectrum evaluated using Eq.~\eqref{eq:Delta_b}. The green shaded region highlights the variation in the baryon power spectrum if $\eta$ goes from $0.1$ to $0.3$. \textbf{Right:} Dimensionless total matter power spectrum for different simulation ICs. The dashed lines are analytical estimate for $\eta=0.3$. These figures show that ICs generated for simulations match with analytical estimate on modes smaller than the Nyquist frequency.} \label{fig:Pk_ngenic_compare}
	\end{figure}

In Fig.~\ref{fig:Pk_ngenic_compare}, we show the matter power spectrum of the ICs generated by {\scshape N-GenIC} for different simulation parameter points described in Table~\ref{table:simparams}. The numerical power spectrum for baryons starts overshooting the analytical estimate near the damping scale. This is because the perturbations start reaching large non-linear values near the damping scales. Consequently, the Zel'dovich approximation used in the generation of initial conditions becomes invalid near the damping scale. This is the reason the Nyquist scales of our simulations do not significantly exceed the damping scale ($\lambda_D$).
	
%	In Fig.~\ref{fig:Pk_ngenic_compare}, we show the matter power spectrum of the ICs generated by {\scshape N-GenIC} for different simulation parameter points described in Table~\ref{table:simparams}. One can see that the numerical power spectrum closely matches the analytical power spectrum. The numerical power spectrum for baryons starts overshooting the analytical estimate near the damping scale. This is because the perturbations start reaching large non-linear values near the damping scales. Consequently, the Zel'dovich approximation \PR{citation?} used in the generation of initial conditions becomes invalid near the damping scale. To minimize the impact of non-linear effects we have chosen box sizes such that the Nyquist scale is only slightly smaller than the damping scale.
	
	The ICs for simulation B were generated with the original {\scshape N-GenIC} algorithm, where particle positions are determined simply by the power spectrum and without generating magnetic fields on the lattice. To account for the influence of PMFs we simply input the PMF-enhanced power spectra for baryons and dark matter, $P(k)=P_{{\rm \Lambda CDM}}(k)+P_{{\rm PMF}}(k)$. This algorithm implicitly enforces baryon and dark matter perturbations to be perfectly correlated with each other. This is because the original {\scshape N-GenIC} algorithm is built for $\Lambda$CDM cosmologies where baryon and dark matter perturbations are sourced from the same stochastic variable (inflationary curvature perturbation) and hence are correlated. However, PMF-induced perturbation is an independent stochastic variable uncorrelated with inflationary curvature perturbation.\footnote{It is possible in some inflationary magnetogenesis theories that PMFs are correlated with curvature perturbations. However, for simplicity, in this study we choose to focus on scenarios where PMFs and curvature perturbations are independent from each other.} The interference between the two sources of perturbations generates an isocurvature between baryon and dark matter perturbations on small scales, i.e. $\delta_{\rm DM}/\delta_{\rm b}$ becomes a stochastic variable. This isocurvature is not captured in the initial conditions for simulation B.

	For simulation C, we modify the {\scshape N-GenIC} code to capture the uncorrelation between PMF-induced perturbations and inflationary perturbations, and hence the isocurvature. Specifically, the density fields are sourced by two independent Gaussian distributions: one originating from $\Lambda$CDM power spectrum and the other from PMF-induced power spectrum (Eq.~\eqref{eq:Delta_b}). The power spectra of dark matter and baryon for simulation C are identical to those of simulation B. However the total matter power spectrum of simulation B and C are different because of the presence of isocurvature between DM and baryon in simulation C.
	
	Our primary method, where particles are displaced according to the Lorentz force, naturally takes into account that PMF-induced perturbations are uncorrelated with $\Lambda$CDM perturbations. Moreover, unlike simulation C, it also takes into account non-gaussianity in matter perturbations and the vortical motion of baryons.\footnote{If PMFs are Gaussian distributed, then the matter perturbations would naturally have non-gaussianities because they are non-linearly related to PMFs, $\delta_{\rm b}\propto \nabla\cdot[(\vec{\nabla}\times\vec{B})\times \vec{B}]$.} However, we find that non-gaussianities and vortical motion play a negligible role in determining baryon fraction of halos.
	
	Earlier studies \cite{Katz:2021,Sanati:2020} that performed similar simulations generated initial conditions similar to our simulation $B$. However, unlike our simulation $B$, they implicitly assumed the equal enhancement to baryon and dark matter perturbations from PMFs, and thus could not capture the enhancement of baryon fraction of high redshift halos. While their initial conditions are less accurate than ours, their simulations include MHD \cite{Katz:2021} or galaxy chemical evolution \cite{Sanati:2020}. In this study, as a first step, we choose to perform a dedicated set of hydrodynamic simulations to focus on the impact of improved initial conditions. 
	
	%Apart from simulations A and D, we evolve our simulations only until redshift 4. We do so because we primarily aim to show the impact of PMFs on high redshift galaxies, where influence of PMFs is most pronounced and astrophysical uncertainties are minimized. 
	
	The following section is organized as follows. In section~\ref{subsec:hmf} we compare the halo mass function obtained from our simulations with $\Lambda$CDM expectations. We show that the Sheth-Tormen analytic formula provides an excellent fit for the numerically obtained halo mass function. In section~\ref{subsec:barfra}, we show the baryon fraction of different halos as a function of their mass and redshift. We find that the presence of isocurvature widens the scatter in the baryon fraction of halos and that the baryon fraction is more enhanced for smaller values of $n_{\rm B}$ and larger values of $\lambda_D$. Finally, in section~\ref{subsec:SF}, we show how the enhanced baryon fraction and matter power spectrum could boost star formation.
	
	\subsection{Halo mass function} \label{subsec:hmf}
	%Dark matter halos are collapsed overdense regions where most galaxies are expected to reside. Halo formation, although gravitationally non-linear in nature, is well understood with the knowledge of linear perturbations. Thus, dark matter halos serve as an essential tool in cosmology, because they enable us to relate linear perturbations with non-linear structures, and consequently distribution of galaxies.  
	
	The halo mass function (HMF) represents the abundance of dark matter halos in a small mass range $[M, M+{\rm d}M]$, per unit volume. In this section, we show that the HMF obtained from simulations is in close agreement with that predicted from theory. Additionally, we discuss the variation of HMF as we change different PMF parameters. 
	
	Theoretical models can directly estimate the HMF from the linear matter power spectrum through the relation
	\begin{equation} \label{eq:HMFdndM}
		\frac{\text{d}n_{\rm h}}{\text{d}M} (M,z) = \frac{\Bar{\rho}_{\rm m}}{M} f(\sigma_{\rm R}) \frac{\text{d}  \sigma_{\rm R}^{-1}}{\text{d} M},
	\end{equation}
	where $\sigma^2_{\rm R}$ is the variance of the smoothed linear overdensity field of the total matter and is determined by
	\begin{equation}
		\sigma^2_{\rm R} \equiv \langle \delta^2 (x,z;{\rm R})\rangle =  \int_0^{\infty}  \left[ \frac{k^3 P(k,z)}{2\pi^2} \right]\Tilde{W}^2(k{\rm R}) \ \frac{\text{d} k}{k}.
	\end{equation}
	Here $\Tilde{W}$ is a Fourier-transformed window function, with the smoothening scale given by the radius ${\rm R} = [3M/4\pi \Bar{\rho}_{\rm m}]^{1/3}$.
	The shape of the differential mass function, $f$ in Eq.~\eqref{eq:HMFdndM}, depends on a particular model of halo collapse or a fitting prescription. In this study, we employ the Sheth-Tormen model \cite{ST1999,SMT2001}, which is determined by the fitting formula of the form
	%\begin{equation}
	%    f_{{\rm S-T}} (\sigma)= A \sqrt{\frac{2a}{\pi}} \left[ 1 + \left( \frac{\sigma^2}{a \delta_c^2}\right)^p % \right] \frac{\delta_c}{\sigma} \exp \left(- \frac{a \delta_c^2}{2 \sigma^2} \right)
	%\end{equation}
	\begin{equation} \label{eq:SMT}
		f_{\text{S-T}} (\Tilde{\nu}) = 2A \left[ 1 + \frac{1}{\Tilde{\nu}{}^{2p}}\right] \frac{\Tilde{\nu}}{\sqrt{2 \pi}} e^{-\Tilde{\nu}{}^2 /2} \, , \quad \Tilde{\nu} = \sqrt{a} \frac{\delta_c}{\sigma_{{\rm R}}},
	\end{equation}
	where $A=0.322$, $a=0.707$, $p=0.3$, and $\delta_c=1.686$.\footnote{Ref.~\cite{Shibusawa:2014fva} analytically finds the collapse threshold in PMF only case (after turning off $\Lambda$CDM perturbations) to be redshift dependent. However, numerically we find that using the constant value of $\delta_c=1.686$ to always provide a good fit for halo abundance.}
    \begin{figure}
	       \includegraphics[width=\linewidth,trim={0 9cm 0 0},clip]{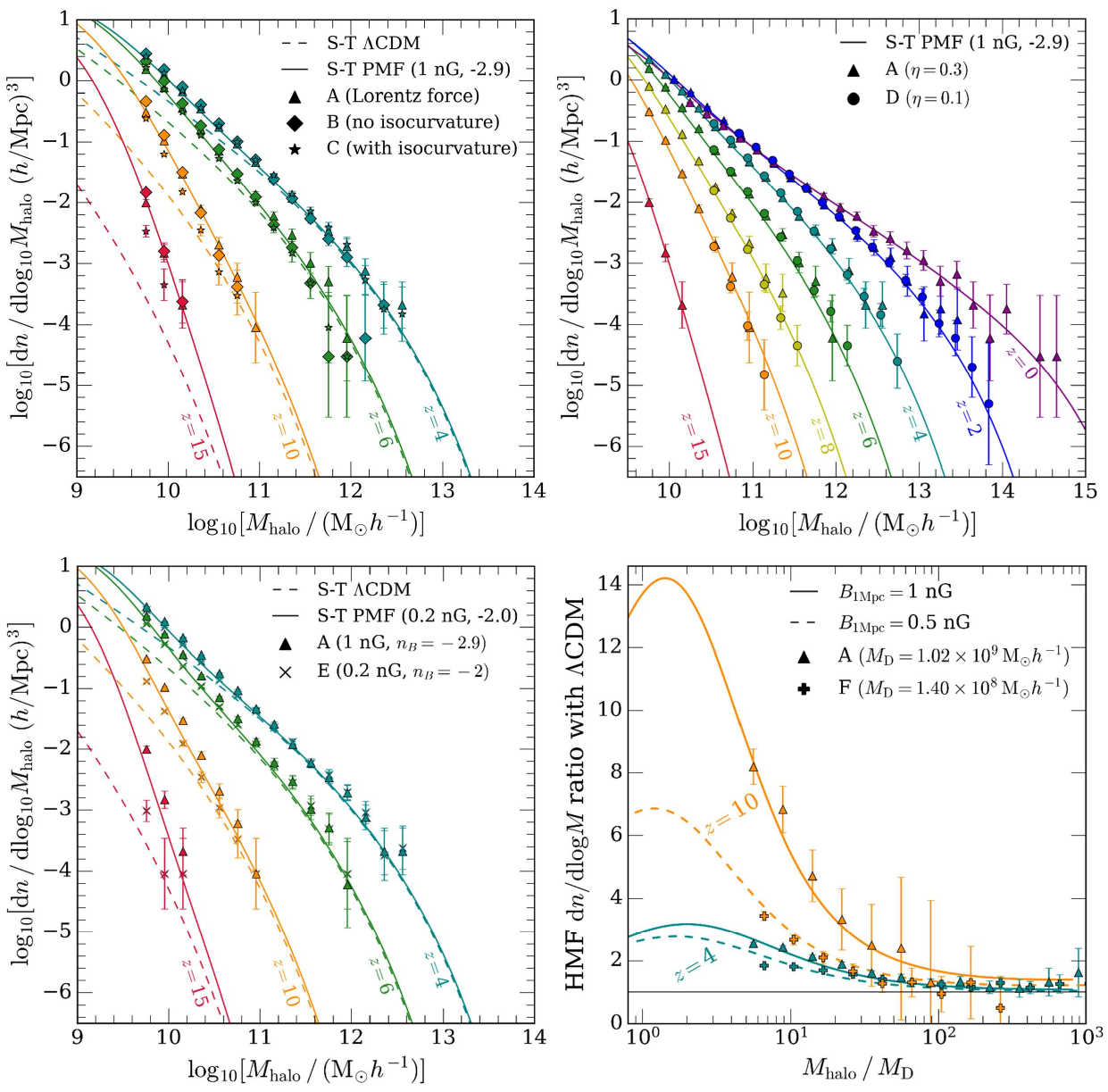}
	
		\caption{{\scshape Halo Mass Function.} Simulation points are shown at the centers of bins with a fixed width $\Delta \log M = 0.2$, count all FoF-identified halos with at least 32 DM particles, and are compared with the Sheth-Tormen theory at several redshifts $z$. Error bars are given by the Poisson shot noise of number of halos in each bin. The S-T lines (dashed for $\Lambda$CDM and solid for PMF in left panels) have different input power spectra, and their parameters are indicated in parentheses for the PMF case. 
			\textbf{Top left}: Simulations A, B and C, with different particle displacement methods in the initial conditions.
			\textbf{Top right}: A and D, with different damping scales given by parameter $\eta$.
			\textbf{Bottom left}: A and E, with different PMF strenghts and spectral indexes, but similar power spectra.
			\textbf{Bottom right}: A and F, shown as ratios of HMFs $F_{{}_{\rm PMF}} / F_{{}_{\rm \Lambda CDM}} $ shifted according to the damping mass $M_{\rm D}$, with $F \equiv {\rm d} n / {\rm d} \log M$; the lines are ratios of S-T fits.}
		 \label{fig:HMF}
    \end{figure} 
	
	In Fig.~\ref{fig:HMF}, we show the Sheth-Tormen HMF at different redshifts. We use the HMF libraries from the package \texttt{Pylians} \cite{pylians} to obtain them.
	The input power spectra were in the $\Lambda$CDM case (dashed lines) obtained directly from \texttt{CLASS}, while in the case of PMF (solid lines) we added analytically computed enhancements, see Eq.~\eqref{eq:Delta_b}. Since PMFs enhance matter power spectrum on small scales we see a greater abundance of low-mass halos compared to $\Lambda$CDM. 
 
    The markers in the plot show the binned HMF obtained from simulations. We identify halos using the Friends-of-Friends (FoF) algorithm \cite{white2002}, with a linking length parameter $b=0.2$, while only considering halos that have at least 32 dark matter particles. The mass of a halo is given by the sum of all components present in the halo, $M_{\rm halo} = M_{\rm DM} + M_{\rm b}$, where $M_{\rm b} = M_{\rm gas} + M_{\ast}$ is the total baryon mass composed of gas and stars. After extracting masses of halos, we construct the binned HMF in the logarithmic space via
    \begin{equation} \label{eq:th_sim_hmf}
		\frac{{\rm d} n_{{\rm h}}}{{\rm d} \log_{10} M} \simeq \frac{\Delta N}{L_{{\rm box}}^3 \Delta \log M},
	\end{equation}
     where $\Delta N$ is the number of halos in each halo mass bin and $\Delta \log M = 0.2$ is the logarithmic bin size. We have tested other values between 0.1 and 0.3, and the results do not depend significantly on this particular choice. The error bars in our HMF plots are related to the Poisson shot noise, for which the best estimate in each bin is $\sigma = 1 / \sqrt{\Delta N}$.
     %\footnote{Note that if $\Delta N$ has a Poisson shot noise error estimate $\sigma_{\Delta N} = \sqrt{\Delta N}$, then, based on the propagation of uncertainty, $F = \log_{10} (\Delta N / const.)$ will have an error estimate given by 
%		\begin{equation*}
%			\sigma_F = \frac{{\rm d} F}{{\rm d} \Delta N} \sigma_{\Delta N} = \frac{1}{ \Delta N \ln (10)} \sqrt{\Delta N} = \frac{1}{ \sqrt{\Delta N} \ln (10)} \,.
%		\end{equation*}
%		However, we choose to show a larger estimate $\sigma = \ln(10) \sigma_F$, because there can be other factors influencing the error, such as binning, choice of $M_{{\rm halo}}$ representing the bin, lack of halos, etc. In addition, the same Poisson error estimate was considered in \cite{Crocce2010}.}
	It is based on the assumption that halos are randomly sampled and form a Poisson realization of the underlying number density \cite{Crocce2010}.
	
	We have not used spherical overdensity (SO) approach in our simulations. This is because some previous studies (e.g., \cite{Jenkins2001, Reed2003}) found that FOF and SO give similar results, but FOF needs less (DM) particles in the low-mass end for determination of the HMF and is computationally cheaper.

	In Fig.~\ref{fig:HMF} we compare our simulations with the Sheth-Tormen theory and find an excellent agreement between the two. As the Sheth-Tormen lines are only a function of the linear matter power spectrum, the figures highlight that HMF is primarily sensitive to the linear matter power spectrum. Thus non-gaussianities in density distribution are unimportant (see appendix~\ref{sec:pdf}). 

    The first panel further supports HMF to only be sensitive to the linear power spectra. Here, the different shaped markers come from simulations that have the same matter power spectrum but in one simulation (A) particles are displaced according to the Lorentz force while in the others (B and C) they are only displaced according to the input power spectrum. 
 
    The second panel shows how the HMF changes for different $\eta$ values. It is apparent that changing $\eta$ and the box size does not affect the HMF on scales larger than the damping scale, $1/k_{\rm D}$. This figure highlights the numerical consistency of our simulations.
	
	%The second and third panels highlight that HMF is primarily sensitive to the linear matter power spectrum and not to the detailed distribution of PMFs. Specifically, both simulations C and E have roughly the same matter power spectrum as simulation A but simulation C does not displace particles according to the Lorentz force while simulation E is for a different PMF configuration (see Fig.~\ref{fig:Pk_an}).

    In the third panel, we vary both $n_{\rm B}$ and $B_{\rm 1Mpc}$ such that the total matter power spectrum for the two simulations is roughly the same (see Fig.~\ref{fig:Pk_ngenic_compare}). Again we see that the HMF for the two simulations are roughly the same as the total matter power spectrad are similar.
    
	Finally, the fourth panel shows the enhancement in HMF for different PMF strengths. The horizontal axis shows the halos mass relative to the damping mass, $M_{\rm D}$, defined as
	\begin{align}
		M_{\rm D}= \frac{4\pi}{3}\rho_{\rm m}(a_0)\times \left(\frac{2\pi}{k_{\rm D}}\right)^3\sim 7.4\times 10^{12}  \left(\frac{h/{\rm Mpc}}{k_{\rm D}}\right)^3\frac{\Omega_{\rm m} h^2}{0.14}\left(\frac{0.678}{h}\right)^2{\rm M}_{\odot}h^{-1}.
	\end{align}
	
	One can see that the shape of the enhancement is very similar and peaks at $M_D$ mass. However, the amplitude of the peak decreases for smaller PMF strengths. The decrease is because smaller PMF strengths enhance the matter power spectrum on smaller scales (see Fig.~\ref{fig:Pk_an}). While the peak of the PMF-induced power spectrum is the same for different PMF strengths, the relative enhancement over $\Lambda$CDM power spectrum decreases because the $\Lambda$CDM power spectrum grows logarithmically on smaller scales.
	
	For PMF strengths chosen in our simulations, $B_{\rm 1Mpc}\leq 1$ nG, there is a significant impact on the abundance of low mass halos, $M<10^{11}\ M_{\odot}$. Consequently, the observed abundance of such halos can be used to put constraints on PMFs that are stronger than the $B_{\rm 1Mpc}\lesssim 1$ nG constraint derived from CMB and Faraday rotation measurements \cite{Planck:2015zrl, PhysRevLett.116.191302}. Indeed Ref.~\cite{Sanati:2020} has used observations of dwarf galaxies to constrain $B_{\rm 1 Mpc}<0.5$ nG for scale invariant PMF. However, we expect these constraints to be modified after taking into account the self-consistent initial conditions we highlight in this study. We leave the detailed evaluation of modified constraints on PMFs from galaxy abundances to future work.
	%(... non-gaussianities in the matter perturbations are negligible).
	
	\subsection{Baryon fraction}
	\label{subsec:barfra}
	In the previous section we saw that at the level of halo mass function, the presence of PMFs was completely parameterised by the matter power spectrum.
	In this section, we show that the baryon fraction of halos can distinguish between scenarios that contain PMFs as opposed to a simple enhancement of the matter power spectrum.
	
	\begin{figure}
		\centering
		\includegraphics[width=0.84\linewidth,trim={0 6.9cm 0 0},clip]{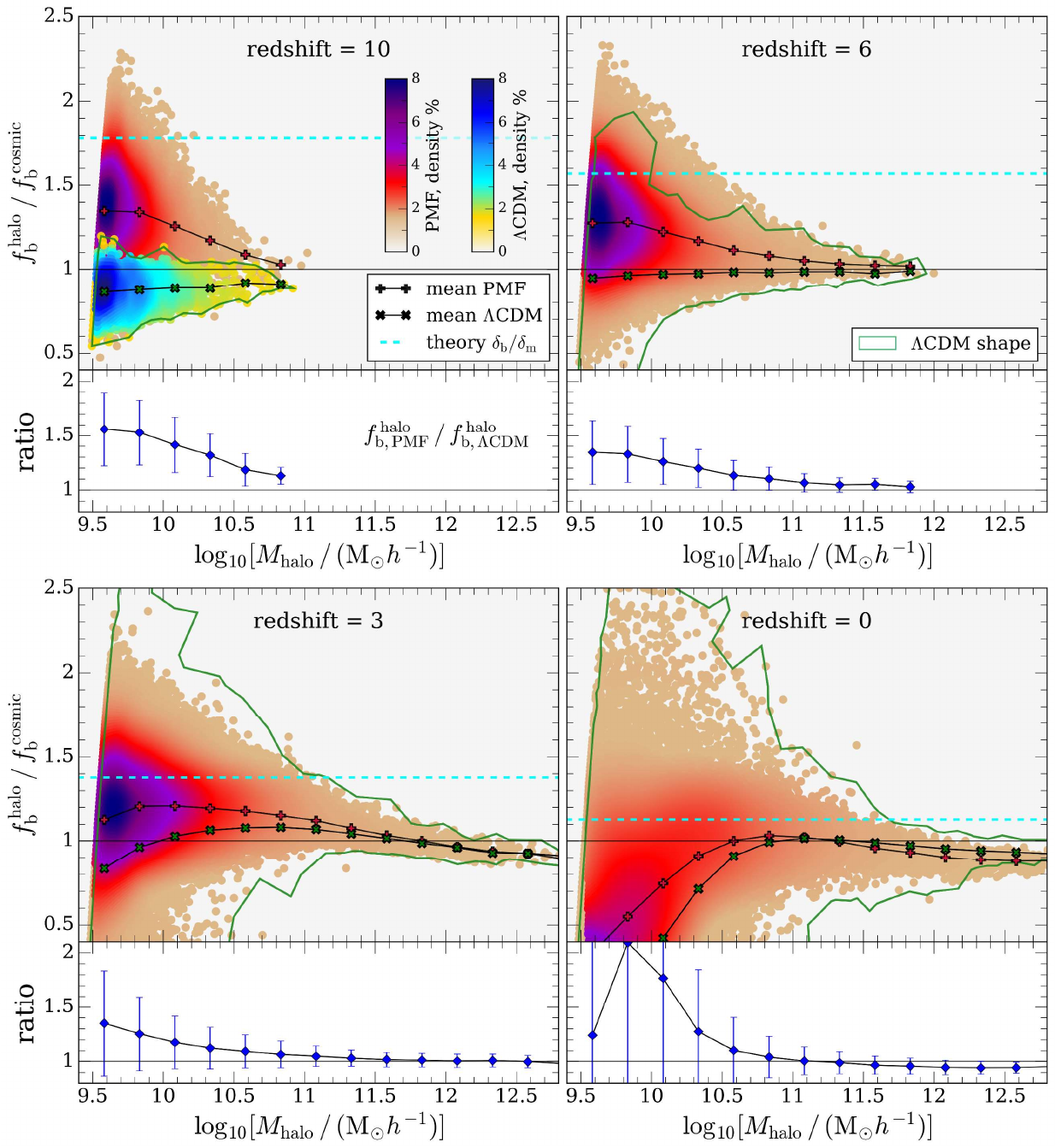}
		\caption{{\scshape Evolution of the baryon fraction in halos.} For the simulation A, the scatter plot shows halo baryon fraction $f_{\rm b}^{\rm halo} = M_{\rm b}/M_{\rm halo}$ with respect to the cosmic mean value and comparison with $\Lambda$CDM-only evolution. Color coding represents the 2D histogram density. Dashed cyan line is a theoretical estimate given by Eq.~\eqref{eq:fb} (see Fig.~\ref{fig:xi}). Error bars in the ratio are computed from standard deviations in each bin. We only show the scatter plot for the $\Lambda$CDM simulation in the first plot. In other plots we simply show the boundary of the $\Lambda$CDM scatter plot as green line for better visualization.
		}
		\label{fig:fb_evolution}
	\end{figure}
	
	The impact of PMFs on the baryon fraction of halos can be clearly seen in Fig.~\ref{fig:fb_evolution}, where we show the scatter plot of all halos in our simulation as a function of halo mass and baryon fraction. The orange-to-purple shaded points are for the simulation with PMFs (simulation A) while the yellow-to-blue shaded points (outlined by a green line) are for $\Lambda$CDM-only simulation. Noticeably, at lower halo masses and higher redshifts, and when PMFs are present, the mean baryon fraction of halos sits precisely between the theoretical estimate $(\delta_{\rm b }/\delta_{\rm m })_{\rm PMF}$ and the mean $\Lambda$CDM value.
    Overall, in the PMF case, halo baryon fractions are larger than the cosmic mean $f_{{\rm b}}^{{\rm cosmic}} = \Omega_{{\rm b}}/\Omega_{{\rm m}}$, while those for $\Lambda$CDM are slightly lower (see also \cite{Crain:2006sb, Castro:2020yes}). In addition, the former is significantly changing with mass, merging with $\Lambda$CDM at the largest masses, reflecting the agreement on large scales, whereas the latter stays relatively constant for high redshifts and considered halo mass interval. The sharp downturn at lower masses can be ascribed mainly to resolution limits \cite{Crain:2006sb}. 
    
    Note that using different approaches to assign gas particles to halo could introduce changes in the baryon fraction of halos. In this study, we only use the FoF algorithm and thus cannot quantify how sensitive the baryon fraction in a halo is to different algorithms. However, the relative enhancement of the baryon fraction in the $\Lambda$CDM+PMF case compared to the $\Lambda$CDM case should not change significantly.
	
	Furthermore, before redshift $\sim 6$, PMFs induce a wider scatter in the baryon fraction compared to the $\Lambda$CDM-only scenario. The large scatter is a result of interference between PMF-induced perturbations and those from inflationary initial conditions. This is more apparent when we rewrite the baryon fraction of a halo, given in Eq.~\eqref{eq:fb}, as
	\begin{align}\label{eq:fb_new}
		\frac{f_{\rm b}}{\bar{f}_{\rm b}}=\frac{\delta_{\rm b}}{\delta_{\rm m}}=\frac{\delta_{\rm b}^{\rm PMF}+\delta_{\rm b}^{\rm \Lambda CDM}}{\delta_{\rm m}^{\rm PMF}+\delta_{\rm m}^{\rm \Lambda CDM}}.
	\end{align}
	Since $\Lambda$CDM and PMF perturbations are uncorrelated and as baryon and dark matter perturbations are amplified differently by PMFs (see Fig.~\ref{fig:Pk_an}), there is an isocurvature component between baryons and dark matter, i.e. $\delta_{\rm b}/\delta_{\rm DM}$ is a stochastic variable. This in turn causes the baryon fraction of halos to also be a stochastic variable.
	
	\begin{figure}
		\centering
		\includegraphics[width=1\linewidth,trim={0 8.4cm 0 0},clip]{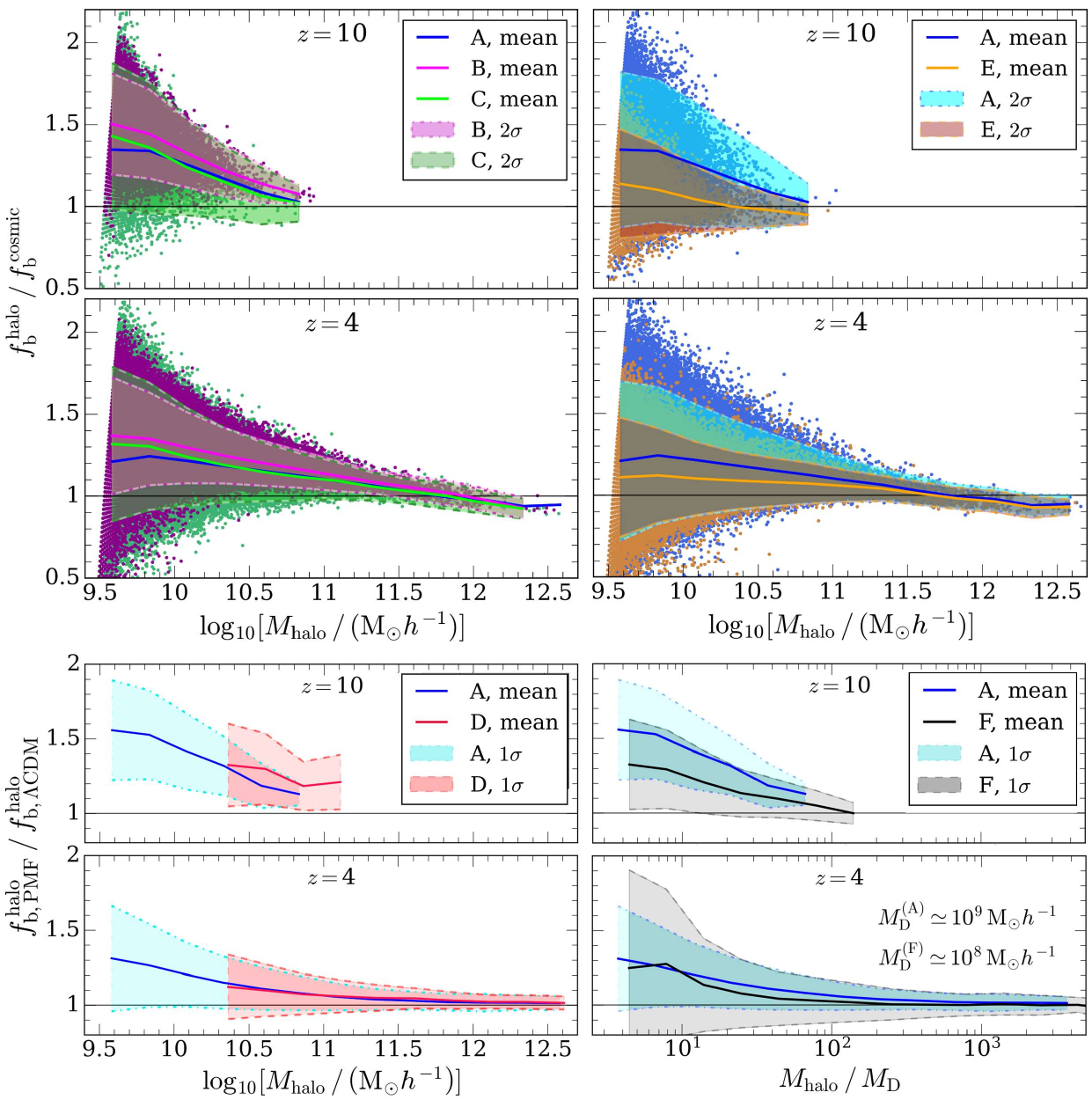}
		\caption{{\scshape Baryon fraction in halos - comparison of different runs.} In the upper part are scatter plots of the halo baryon fraction in units of the cosmic mean value. In the lower part are ratios with $\Lambda$CDM simulations data, for corresponding box sizes. Means and error bands ($1\sigma$ or $2\sigma$) assume approximate Gaussian statistics in each bin. {\bf Top left:} Simulations A, B and C, with scatter points and $2\sigma$ regions shown only for B and C. 
        {\bf Top right:} A (1 nG) and E (0.2 nG), with similar power spectra.
        {\bf Bottom left:} A and D, with different damping scales quantified by $\eta=0.1$ and $\eta=0.3$, respectively.
        {\bf Bottom right:} A and F (0.5 nG), with the x-axis shifted according to the damping mass $M_{\rm D}$.}
		\label{fig:fbABCDEF}
	\end{figure}
	
	We have verified that the large scatter in the baryon fraction for the PMF scenario as seen in Fig.~\ref{fig:fbABCDEF} is due to $\Lambda$CDM and PMF perturbations being uncorrelated. Specifically, in the top left panel, we show the mean and variance of the baryon fraction as a function of halo mass for simulations A, B and C. One can see that simulations A and C produce almost identical results, while simulation B has a narrower scatter (reflected also by a narrower $2\sigma$ band). The narrow scatter is because the PMF and $\Lambda$CDM perturbations are perfectly correlated and hence the baryon fraction in halos (Eq.~\eqref{eq:fb_new}) is not a stochastic variable. Whereas simulation C takes into account the isocurvature between dark matter and baryons.
	
	The isocurvature between dark matter and baryon is also the reason behind the mean of the baryon fraction scatter in Fig.~\ref{fig:fb_evolution} to be smaller than the analytical estimate (cyan dashed line) calculated in section~\ref{sec:grow}. Specifically, in section~\ref{sec:grow} we estimated the baryon fraction by only focusing on PMFs and neglecting contribution from inflationary initial conditions. This amounted to setting $\Lambda$CDM marked perturbations in Eq.~\eqref{eq:fb_new} to zero. Once we include the contribution from $\Lambda$CDM marked perturbations we decrease the baryon fraction because the denominator in Eq.~\eqref{eq:fb_new} increases more relative to the numerator.
	
    The slope of the cutoff of baryon fraction scatter, as observed on the left side of top panels in Fig.~\ref{fig:fbABCDEF}, is a direct consequence of defining halos with at least 32 dark matter particles. More specifically, the halo baryon fraction from simulations reads
    \begin{equation} \label{eq:fbhalo}
        f_{\rm b}^{\rm halo} = \frac{M_{\rm b}}{M_{\rm b} + M_{\rm DM}} = \left[1 + \frac{\Omega_{\rm DM}}{\Omega_{\rm b}} \frac{N_{\rm DM}}{N_{\rm b}} \right]^{-1} \,,
    \end{equation}
    where we used the fact that each species' mass in the halo is given by $M_s =N_s m_s = N_s \Omega_s \rho_{\rm cr} (L_{\rm box}/N_{\rm mesh})^3$, where $N_s$ is the number of particles in the halo of species $s$. Thus, the initial slope of the baryon fraction scatter is given simply by having $N_{\rm DM} = 32$ and adding baryons one by one, until it reaches a certain maximum value. Unsurprisingly, the theoretical formula for baryon fraction \eqref{eq:fb} can be written in the same way as \eqref{eq:fbhalo}, by interchanging $N_s \leftrightarrow \delta_s$.
    
    %The maximum value of baryon fraction scatter is generated by the smallest halos because the PMF-induced power spectrum is most enhanced compared to $\Lambda$CDM power spectrum at the smallest scales. While the accretion of matter brings the mean baryon fraction of halos to cosmic levels with time, the maximum scatter of baryon fraction roughly stays the same. We hypothesize that the non-linear interactions between halos, such as mergers, also lead to scatter and hence counter the decrease caused by accretion.
    
    %This maximum is not changing until very low redshifts because there are two things balancing each other: the ratio $\delta_{\rm DM}/\delta_{\rm b}$ is growing, which would lower the maximum values at a given mass, but there are also interactions of halos at play -- similarly as $\Lambda$CDM scatter gets wider, the PMF scatter is undergoing the same widening (see bottom right panel of Fig.~\ref{fig:fb_evolution}).
 
    Finally, it is worth noticing that at lower redshifts, the scatter in the baryon fraction for the PMF scenario overlaps more with the $\Lambda$CDM scenario. There are two reasons for this. First, as halos accrete more matter, their baryon fraction asymptotes to the cosmic average and hence the mean scatter of both $\Lambda$CDM and PMF scenario eventually asymptotes to the same values. Second, accretion and mergers of halos tend to increase scatter with time. For the PMF scenario, the halos are formed with a large scatter in the baryon fraction due to the presence of isocurvature. Hence, we believe the further enhancement of scatter by mergers/accretion is minimal. In contrast, in the $\Lambda$CDM scenario, the halos are initially formed with minimal scatter and accretion/mergers play a major role in the evolution of baryon scatter.
	
	In Fig.~\ref{fig:fbABCDEF} we also show how baryon fraction scatter changes as we change PMF parameters. Increasing $n_{\rm B}$ results in a reduction of baryon scatter (top right panel of Fig.~\ref{fig:fbABCDEF}). We postulate that this is because the range of scales where PMFs enhance the matter power spectrum over the $\Lambda$CDM one is reduced as $n_{\rm B}$ increases. Thus the baryon fraction quickly asymptotes to the $\Lambda$CDM value as halo mass increases.
 
    As we change $\eta$ from 0.3 to 0.1 (bottom left panel of Fig.~\ref{fig:fbABCDEF}), the baryon fraction scatter almost stays the same for large masses. This is expected as $\eta$ only affects the location of the damping scale and thus properties of large massive halos should remain unchanged. 
	
	Finally, reduction in the strength of PMF, $B_{\rm 1Mpc}$, causes the baryon fraction enhancement to move to smaller masses because PMF-enhanced power spectra move to smaller scales (see Fig.~\ref{fig:Pk_an}). Thus, when we plot the enhancement of baryon fraction as a function of $M_{\rm halo}/M_{\rm D}$ in Fig.~\ref{fig:fbABCDEF}, we see a similar shape for both A and F simulation. However, simulation F has a slightly smaller enhancement in $f_{\rm b}$ because, on small scales, the logarithmic growth of $\Lambda$CDM power spectrum increases the contribution of $\Lambda$CDM $f_{\rm b}$ values in the halos.

    The baryon fraction shown in Figs.~\ref{fig:fb_evolution}-\ref{fig:fbABCDEF} could be suppressed by baryonic feedback processes (AGNs, winds, UV/X-ray background), which are not accurately modelled in our simulations. Specifically, our simulation has no feedback from AGNs and winds and we believe these effects can dramatically affect baryon fraction in halos \cite{Ayromlou:2022jma, 2022MNRAS.516..883S}. 
    
    Our simulation does model UV background (UVB) based on Ref.~\cite{Puchwein:2018arm}. The implemented UV background is consistent with observational data on QSO/galaxy luminosity functions, thermal history of the IGM, timing of reionization, and high-redshift Lyman-$\alpha$ forest data. However, this UVB model not take into account explicitly the star formation rate from the simulation. Since star formation rate for the case with PMFs is expected to be much larger than the case for $\Lambda$CDM at high redshifts (see section~\ref{subsec:SF}), one could expect a much larger UVB in the case of PMFs. In Ref.~\cite{Ayromlou:2022jma}, the primary impact of the star-formation UV feedback is to suppress the accretion of gas for $M_{\rm halo}<10^{10} {\rm M}_{\odot}$.  Since in the case of PMF-induced structure formation, halos are formed with large baryon fraction before star-formation, the local UVB-related suppression of accretion should have minimal impact on the final baryon fraction.

	\subsection{Analyzing impact on star formation}
	\label{subsec:SF}
	In the previous section, we saw that PMFs can significantly enhance the baryon fraction of high redshift halos. However, directly observing the baryon fraction of these halos poses significant challenges. Instead, a more promising observable is the stellar mass of halos. In this section, we show how PMFs might enhance the stellar mass of high redshift halos. Since our simulations do not incorporate MHD, the results of this section should not be taken as a final verdict on PMF's impact on star formation.
	
	In our simulations, the star formation is based on the multi-phase model of Springel \& Hernquist \cite{springel03}.  
%     This model captures in an effective way the multiphase nature of star-forming regions on sub-galactic scales that are typically not resolved in cosmological simulations. By assigning a hot and cold phase to each gas particle, and with the inclusion of gas radiative heating and cooling, it simulates more physically the growth of cold clouds embedded in an ambient hot medium and returns galaxies with realistic physical properties. Our implementation includes thermal feedback from supernovae, but lacks winds and outflows.
The model assumes that the gas is composed of a hot ($T \geq 10^5$ K) and a cold component ($T \leq 10^5$ K). Cold clouds ($T \simeq 10^3 $K) are formed due to thermal instability between the two gas phases. The clouds are allowed to grow due to the radiative cooling of the hot phase. Our simulations form stars within cold clouds, at locations where overdensity exceeds $\delta=10^3$. A 10\% fraction of formed stars is set to be short-lived and explode as supernovae, which heats the ambient hot gas and leads to the evaporation of cold clouds. Such a stellar feedback provides a self-regulatory star formation. This is the only feedback mechanism included in our simulations. The introduction of other feedback processes such as AGNs and galactic winds could further suppress star formation as they reduce the amount of gas available for star formation or eject the gas into halo outskirts.

%The above model leads to an overall overproduction of stars. This is because the radiative cooling of the hot gas into cold clouds more than compensates for the reduction of cold clouds from supernovas.
% However, there will be an overall overproduction of stars, because the radiative cooling is very efficient at high redshifts and small halos, and baryons have no way to escape out of the gravitational potential of their galaxies, once they fall into them due to cooling. Thus, 
%For a more realistic description, one would have to introduce also the galactic winds, which either locally reduce the gas available for star formation or eject the gas into halo outskirts. In our numerical implementation, these star-formation suppression mechanisms are not included. Thus, we emphasize that the star formation results below should be understood as a comparison of simulations with PMFs and without.
%Nevertheless, we emphasize that the star formation results below should be understood as comparison of simulations with PMFs and without. In general, the winds reduce the star formation in low-mass halos the most.
     
From the perspective of PMF-induced structure formation, a relevant feature of the above prescription is that the mass fraction of cold clouds grows monotonically for very high baryonic overdensities. Since with PMFs, we can reach higher baryon overdensities in halos compared to $\Lambda$CDM, we expect a larger production of star-forming clouds and thus elevated star formation.

\begin{figure}
		\centering
		\includegraphics[width=0.9\linewidth]{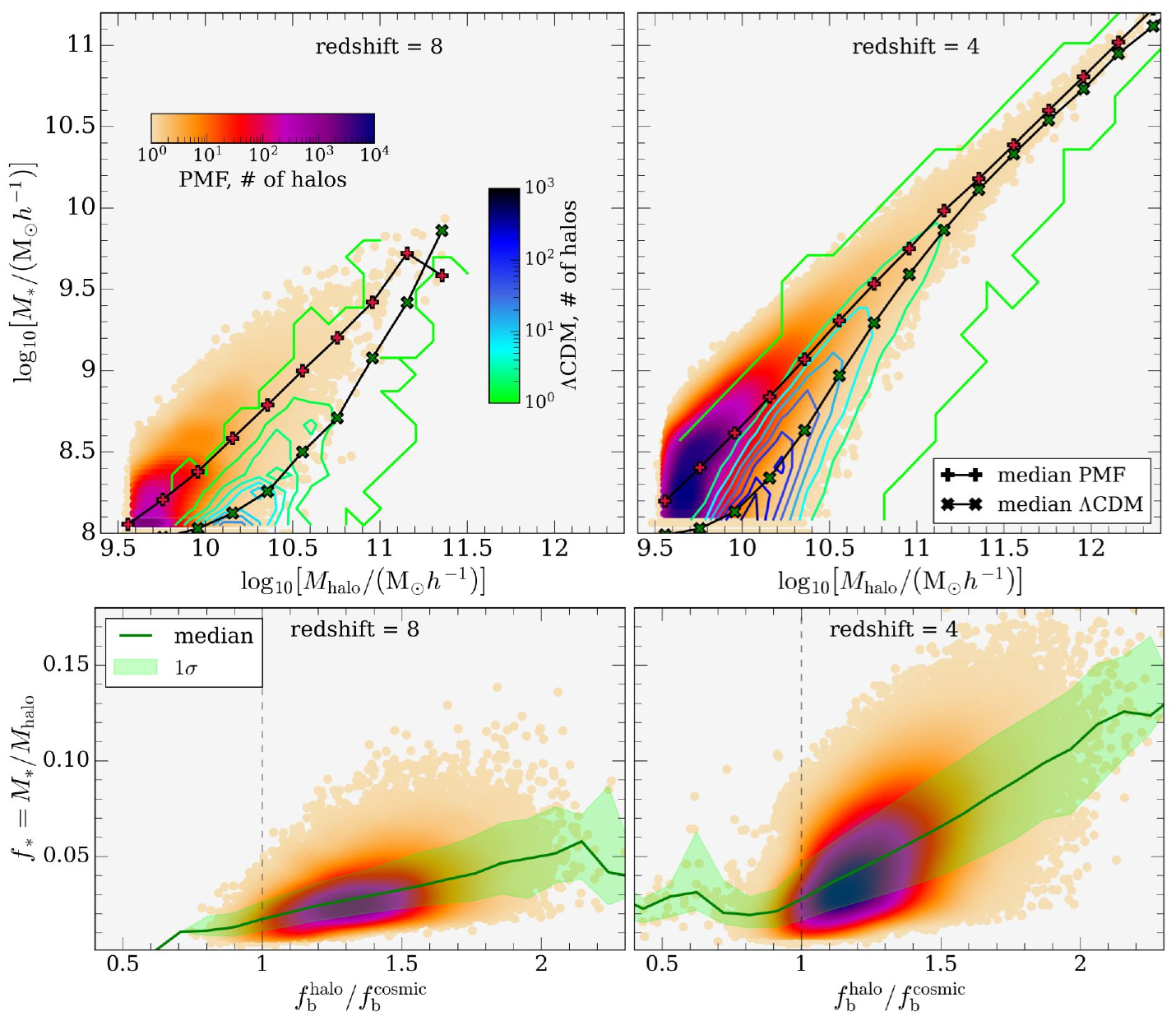}
		\caption{{\scshape Stellar mass and star fraction in halos.} For the simulation A, the scatter plots show halos which contain at least 1 star particle. {\bf Top:} Stellar vs.~halo mass in the log-log plane. $\Lambda$CDM is shown with colored contours. {\bf Bottom:} Relationship between $f_{*}$ and $f_{\rm b}^{\rm halo}$ for the PMF simulation. The 1-sigma error band around the median is given by the 16/84-th percentiles in each $f_{\rm b}$ bin.}
		\label{fig:Mstar_fstar}
	\end{figure}
 
In the upper part of Fig.~\ref{fig:Mstar_fstar}, we show a scatter plot of halos as a function of their stellar mass and halo mass. One can see that the stellar mass in halos is significantly enhanced compared to the $\Lambda$CDM simulation for $M_{\rm halo}<10^{11}\ {\rm M}_{\odot}$.
%As we use the same prescription of star formation for both $\Lambda$CDM only runs and ones with PMFs, the difference in star formation is solely due to the enhancement in baryon fraction and matter power spectrum. Star formation is primarily enhanced compared to $\Lambda$CDM for halos with masses smaller than $10^{11}\ {\rm M}_{\odot}$. These correspond to scales where PMFs enhance the matter power spectrum.
To estimate how much of the enhancement of stellar mass comes directly from higher baryon fraction, in the bottom panel of the Fig.~\ref{fig:Mstar_fstar} the scatter plot is shown as a function of $f_*\equiv M_{*}/M_{\rm halo}$ and baryon fraction of halos. One can see that there is a direct correlation between the two, however, the correlation is weak as there is a significant scatter around the median. Additionally, we also find that the correlation between $f_*$ and $f_{\rm b}$ increases with time. The increase in correlation could be a consequence of larger time allowing conversion of more gas to stars, compared to the standard case. If this hypothesis is true, one would have an increasing value of the gas to star formation efficiency, $\epsilon\equiv M_*/M_{\rm b}$, with time.

\begin{figure}
		\centering
		\includegraphics[width=1\linewidth,trim={10.5cm 6.1cm 0 6.1cm},clip]{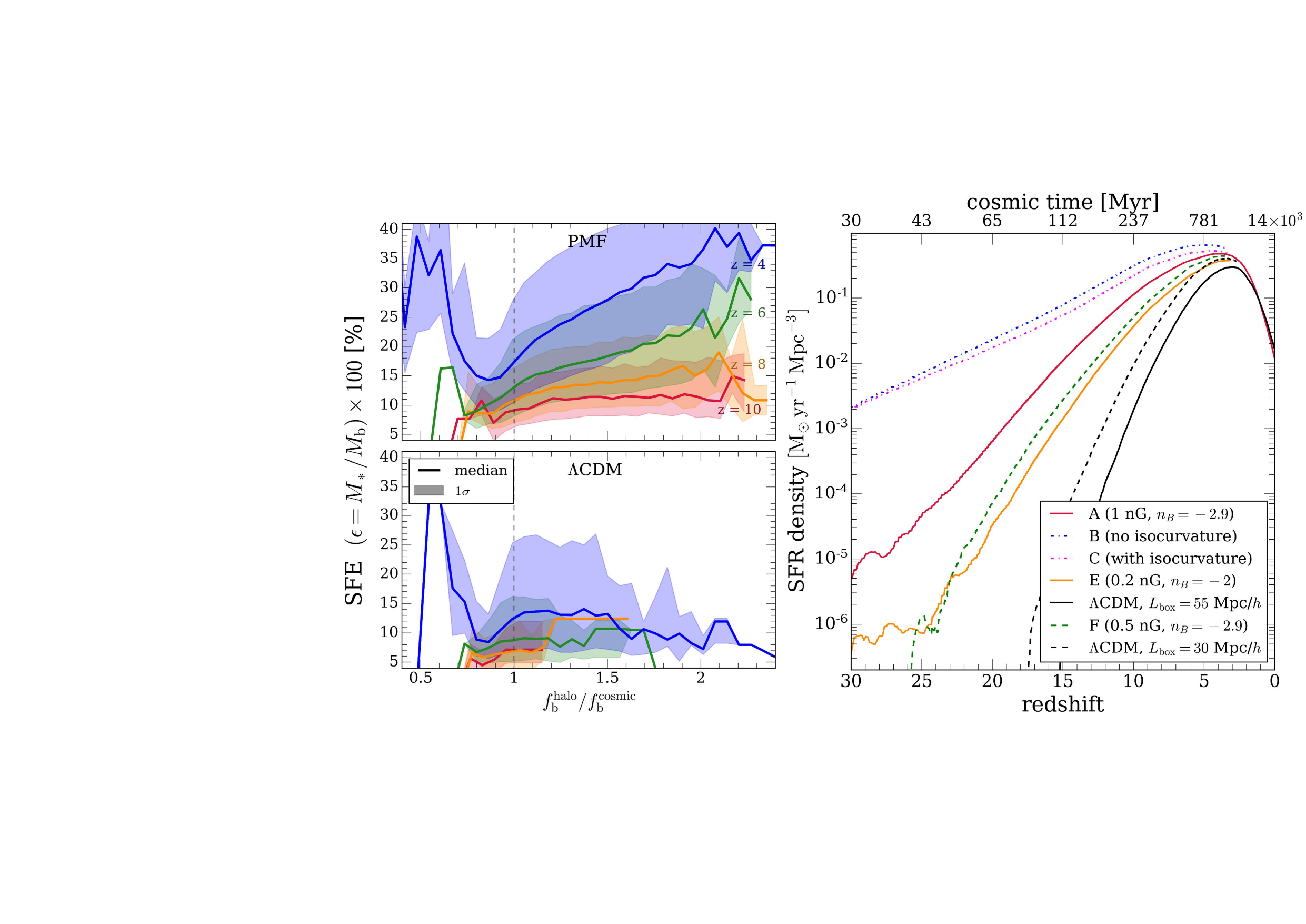}
		\caption{{\scshape Star formation efficiency and star formation rate.} {\bf Left:} Evolution of SFE, expressed as the efficiency of gas conversion into stars $\epsilon$ in relation to halo baryon fraction, for the simulation A (top) and $\Lambda$CDM (bottom). {\bf Right:} Density of the total SFR (integrated over entire volume) for each simulation (see Table \ref{table:simparams}).}
		\label{fig:SFR_SFE}
	\end{figure}

In the left panel of Fig.~\ref{fig:SFR_SFE} we show the mean and 1$\sigma$ scatter of halos as a function of $\epsilon$ and $f_{\rm b}$ for different redshifts. As expected, the median value of $\epsilon$ increases with time for each $f_{\rm b}$ bin. However, at larger $f_{\rm b}$ values one observes a larger growth in $\epsilon$. This dependence of $\epsilon$ on $f_{\rm b}$ at lower redshifts suggests a non-trivial dependence of star formation on baryon fraction, which is absent in $\Lambda$CDM simulation. As $\Lambda$CDM simulations obtain large $f_{\rm b}$ values through accretion and mergers, whereas PMF simulations have an enhanced $f_b$ because of the initial condition itself, the dependency of $\epsilon$ on $f_b$ seen in PMF simulation likely corresponds to the historical evolution of $f_b$.

We also find PMFs to boost the star formation rate (SFR) density compared to $\Lambda$CDM as seen in the right panel of Fig.~\ref{fig:SFR_SFE}. The enhancement at high redshifts, $z>10$, is primarily because halos collapse earlier due to the boosted power spectrum. Consequently, star formation begins much more in advance compared to $\Lambda$CDM. The suppression of SFR after $z<6$, irrespective of the PMF strengths, is due to our simulations using a reionization model that is independent of star formation history \cite{Puchwein:2018arm}.

Interestingly, the star formation rate in simulations B and C is much higher than the rate in simulation A, even though all these simulations have similar halo mass functions and baryon fractions. This contrast arises from the absence of vortical motion in simulations B and C, where particles are not displaced according to the Lorentz force, as opposed to simulation A. Consequently, we hypothesize that the augmented angular momentum of halos induced by vortical motions acts to suppress star formation. 

Note that our star formation prescription does not model feedback from winds or active galactic nuclei. For $\Lambda$CDM simulations, without such feedback mechanisms, it was found \cite{springel03} that there is an overall overproduction of stars when compared to local disk galaxies. The main reason for such an overproduction is that radiative cooling is very efficient for high redshift and low-mass halos, with no mechanism for cold gas to escape from the gravitational potential of star-forming regions. In our PMF simulations, we should expect the same overproduction, which could, however, be even higher. This is because baryon overdensity enhancement is most pronounced for high redshifts and small halos.
The inclusion of AGN/wind feedback might non-trivially change the behaviour of star formation rate and needs to be accounted for before connecting simulations with observations \cite{2024arXiv240407252S}. 

Finally, it is well-known that magnetic fields provide additional support against the collapse of molecular clouds into stars and play a role in the physics of the interstellar medium. Therefore, the evolution of PMFs, all the way to galaxy formation, should bring additional insights into our star formation results. It is worth mentioning that values of PMF strengths less than 1 nG were found to minimally impact the star formation histories \cite{Marinacci:2015ria}. However, this study assumes a uniform magnetic field configuration in the initial conditions. A stochastically distributed primordial magnetic field that is correlated with the displacement of particles in the initial conditions might change the result non-trivially.

	\section{Summary and Conclusions}\label{sec:con}
	In this work, we carefully delved into how primordial magnetic fields (PMFs) influence baryon and dark matter perturbations. We have pointed out three important aspects that have not been scrutinized thoroughly in previous research.
	
	First, we found that PMFs can produce halos at high redshifts ($z>4$) which have a baryon fraction several times larger than the cosmic average. This happens because PMFs directly boost baryon density perturbations, while dark matter perturbations only get a boost from the gravitational pull of baryon perturbations. As halos accrete surrounding matter, their baryon fraction eventually levels off to the cosmic average at lower redshifts.
	
	Secondly, PMFs introduce isocurvature between baryon and dark matter perturbations. This happens because the perturbations induced by PMFs are uncorrelated with the perturbations sourced by inflationary initial conditions in the standard cosmological model. The lack of correlation, combined with the asymmetric effect of PMFs on baryon and dark matter perturbations, leads to isocurvature. We show that this isocurvature increases the scatter in baryon fraction among high-redshift halos.
	
	Lastly, we have shed light on the complexity of figuring out the magnetic damping scale (or magnetic Jeans scale). This scale sets the limit beyond which baryons' backreaction on magnetic fields suppresses the baryon power spectrum. Current estimates of the magnetic Jeans scale rely on simple calculations, which could overestimate the peak of the baryon power spectrum by orders of magnitude. To address this, we introduce a new variable, $\eta$,  in order to parameterize the uncertainty in the magnetic damping scale and discuss our results for different $\eta$ values. 
	
	To confirm the influence of PMFs on baryon fraction, we ran hydrodynamical simulations using \texttt{P-Gadget-3} code. It is important to note that our simulations did not include magnetohydrodynamics (MHD), and we only considered the effect of PMFs through the initial displacement field of particles. Since we ran the simulations after gravity overcame the Lorentz force from PMFs, the absence of MHD should have a minimal impact on the baryon fraction of halos in our simulations.
	
	For our simulations, we selected PMF strengths permitted by CMB and Faraday rotation measurements, specifically $B_{\rm 1Mpc} \lesssim 1$ nG. However, these parameter points may already be constrained by observations of Lyman-alpha, reionization optical depth, or galaxy abundances. Previous studies have derived constraints on PMFs from these measurements, but they did not account for several critical effects of PMFs on density perturbations that we have highlighted in this paper. We defer the task of estimating these modified constraints on PMFs to future research.
    
    Through our simulations, we found that non-gaussianities and vorticity of PMF-induced perturbations had negligible impact on halo mass function and baryon fraction of halos but still may play a non-trivial role in star formation. In particular, we ran two simulations: one where particles in the initial conditions were displaced along Lorentz force field lines, and a second where particles were displaced with a random Gaussian distribution determined by the power spectrum. Both simulations were identical at the level of matter power spectrum but the second simulation could not capture the non-guassianity and vorticity present in the first one. However, the two simulations were found to provide identical halo mass function and baryon fraction statistics. The star formation rate was found to be suppressed in the simulation with Lorentz force-displacement and we hypothesize it could be due to the larger angular momentum of halos. Moreover, we confirmed that the Sheth-Tormen formalism for calculating halo mass functions remains applicable even for the PMF-enhanced matter power spectrum.
	
	Next, our simulations demonstrated that the baryon fraction of halos could be twice the cosmic average at redshifts of $z\sim 10$. This enhancement is most prominent for halos around the magnetic damping scale and levels off to the cosmic average for larger masses. Changing the strength of PMFs increases the magnetic damping scale, shifting the enhancement of the baryon fraction to larger halos. Altering the spectral slope of PMFs while keeping the damping scale fixed reduces the enhancement in the baryon fraction. This is because a larger spectral slope concentrates the impact on density perturbations to a smaller range of scales near the damping scale. Thus for halos on scales larger than the damping scales, i.e. scales where we trust the simulation results, we observe smaller enhancement in baryon fraction.
	
    A higher baryon fraction as well as an enhanced small-scale power spectrum can boost star formation. Using the multi-phase model for star formation \cite{springel03}, we find that PMFs primarily enhance star formation in halos corresponding to the scales where PMFs enhance the matter power spectrum. The gas-to-star conversion efficiency, $\epsilon$, was boosted on average by a factor of 2 to 3 compared to $\Lambda$CDM. Moreover, $\epsilon$ was found to develop a positive correlation with the baryon fraction with time, while such a correlation was absent in $\Lambda$CDM simulations.

    Our results on star formation could be significantly modified after including more physical effects due to magnetic fields in our simulations. These include: 1) using MHD; 2) a star formation prescription that takes into account magnetic fields, 3) a self-consistent treatment of heating for the intergalactic medium from primordial magnetic field's dissipation; 4) appropriately calibrated  galaxy formation models which include feedback from Supernovae winds and/or Active Galactic Nuclei. 

    In contrast, the baryon fraction of halos should be largely insensitive to the above-mentioned effects, except for the feedback from black holes and supernovas. However, these feedback processes typically reduce the baryon fraction in halos.  
    %Moreover, the feedback processes in standard cosmology are typically unimportant at high redshifts, $z>6$. 
    The only processes that can enhance baryon fraction are non-linear interactions between halos and the existence of primordial magnetic fields. As in standard cosmology, the interaction between halos is not expected to produce $f_{\rm b}>1.5$ at high redshifts ($z>8$), a detection of such enhanced baryon fraction values in high-redshift galaxies could provide a smoking-gun signal for the primordial nature of magnetic fields.

    It is intriguing that recent observational campaigns have discovered: 1) a population of high-redshift and very luminous galaxies whose physical properties might be difficult to reconcile within the standard structure formation model (e.g. \cite{labbe23,boylan}); 2) detection of a very massive black hole at $z\sim 11$ which again might hint at an accelerated phase of star formation (and accretion from the surrounding medium) \cite{maiolino23}; 3) a consistent picture of the rapid growth of black holes and galaxies at $z>6$ from ALMA radio observations of quasar (QSO) environments in which star formation is found to be very high \cite{tripodi}; 4) bursty star-forming galaxies at $z=[4,6]$ and signatures of large galaxy-galaxy interactions \cite{asada}; 5) relatively large gas fractions in $z>6$ QSO host galaxies \cite{neeleman21}.
    All these observational results are likely to be impacted at some level by PMFs and could potentially return galaxies in better agreement with the data. 
    Thus, it is clear that in the next few years, thanks to the combined efforts of JWST and ALMA, the baryon census in high redshift galaxies will become more complete and accurate providing a unique data set to test PMF models.

	\appendix
	\section{Ionization fraction and ambipolar diffusion}
	\label{sec:AmbDiff}
 
 When a magnetic field is present in a partially ionized gas, the Lorentz force acts directly only on charged particles and, thus, induces a relative motion between the charged and a neutral component. This ambipolar drift, also known as ambipolar diffusion (AD), is an important part of many astrophysical processes. However, frequent Coulomb collisions between ions and polarized neutrals tend to suppress the relative drift motions of the two constituents, and in such cases the AD may be neglected. In other words, neutral atoms and molecules feel the Lorentz force through scattering with ionized ones. On the other hand, if the number of ions is insufficient, they cannot impose their momentum onto the neutrals, and the latter would be decoupled from ions. Here we seek to estimate what is the minimal ionization fraction needed, so that the neutral component moves together with ions, while keeping the relative drift negligible. Our focus is set on redshifts around $z \simeq 100$. Related discussions can be found in \cite{Sethi:2004pe} and \cite{Banerjee:2004df}.

 We denote the physical (non-comoving) number densities, mass densities and velocities of ions and neutrals as $n_{\rm i}$, $\rho_{\rm i}$, $\vec{v}_{\rm i}$ and $n_{\rm n}$, $\rho_{\rm n}$, $\vec{v}_{\rm n}$, respectively, with the total baryon density and mean fluid velocity being given by $\rho_{\rm i} + \rho_{\rm n} = \rho_{\rm b}$ and $\Vec{v}_{\rm b} = (\rho_{\rm i}\Vec{v}_{\rm i} + \rho_{\rm n} \Vec{v}_{\rm n})/\rho_{\rm b}$. The contribution of electrons may be neglected in the following considerations, owing to their low mass. The densities represent mean cosmological values at a given time. The ionization fraction is $X_{\rm i} \equiv n_{\rm i}/(n_{\rm i}+n_{\rm n})$.
 We have the following MHD equations 
 \begin{align}
        \rho_{\rm i} \left[ \frac{\partial}{\partial t} + H + \frac{\Vec{v}_{\rm i} \cdot \nabla}{a}\right] \Vec{v}_{\rm i}  &= \frac{1}{a^5} \frac{(\nabla \times \Vec{B}) \times \Vec{B}}{4\pi} -  \rho_{\rm i} \alpha_{\rm in} \left(
            \Vec{v}_{\rm i}  - \Vec{v}_{\rm n}\right) \,, \label{eq:AD_ions}\\ 
         \rho_{\rm n}\left[ \frac{\partial}{\partial t} + H + \frac{\Vec{v}_{\rm n} \cdot \nabla}{a}\right] \Vec{v}_{\rm n}  &= -  \rho_{\rm n} \alpha_{\rm ni} \left(
            \Vec{v}_{\rm n}  - \Vec{v}_{\rm i}\right) \,, \label{eq:AD_neutrals} \\ 
            \frac{\partial \Vec{B}}{\partial t} &= \frac{1}{a} \nabla \times \left( \Vec{v}_{\rm i} \times \Vec{B} \right) \,,\label{eq:AD_induction}
 \end{align}
 where $\alpha_{\rm in} = \rho_n \langle \sigma w\rangle_{\rm in}/(m_i + m_n)$ is the mean collision rate of ion-neutral Coulomb interactions, and it holds  $\rho_{\rm n}\alpha_{\rm ni} = \rho_{\rm i}\alpha_{\rm in}$. We remind that $a$ is the scale factor, $\Vec{B}$ is a comoving field and $\nabla$ are comoving derivatives. 
 
 When the ions and neutrals are `tightly coupled', the collisions are highly frequent, which translates to the mean collision time, $\tau_{\rm in} = \alpha_{\rm in}^{-1} $, being very short compared to a characteristic time scale emerging from the l.h.s. of Eq.~\eqref{eq:AD_ions}. In such a case we can neglect the whole l.h.s. in the mentioned equation and find for the drift velocity 
 \begin{equation}\label{eq:drift_vel}
      \Vec{v}_{\rm drift} \equiv \Vec{v}_{\rm i} - \Vec{v}_{\rm n} \approx \frac{1}{a^5}\frac{(\nabla \times \Vec{B})\times \Vec{B}}{4\pi  \rho_{\rm i} \alpha_{\rm in}} \,.
 \end{equation} 
 
 In the tight coupling regime and for a poorly ionized gas ($\rho_{\rm n} \approx \rho_{\rm b}$ and $\Vec{v}_{\rm n} \approx \Vec{v}_{\rm b}$), one may treat the baryonic gas as a single fluid with $\rho_{\rm b}$ and $\Vec{v}_{\rm b}$, on which the Lorentz force acts. This can be seen after substituting for drift velocity \eqref{eq:drift_vel} into Eq.~\eqref{eq:AD_neutrals} for neutrals 
 \begin{equation} \label{eq:Lor_for_neutrals}
       \left[ \frac{\partial}{\partial t} + H + \frac{\Vec{v}_{\rm b} \cdot \nabla}{a}\right] \Vec{v}_{\rm b}  = \frac{1}{a^5} \frac{(\nabla \times \Vec{B}) \times \Vec{B}}{4\pi \rho_{\rm b}} \,.
 \end{equation}
%It can be seen from this equation why neutral particles, and thus the whole fluid, feel the Lorentz force.

 Our tight coupling approximation is valid as long as $v_{\rm drift} \ll v_{\rm b}$. One can make an approximate check for the validity of tight coupling by requiring the average values of $v_{\rm drift}$ and $v$ to satisfy the mentioned inequality. Using Eq.~\eqref{eq:drift_vel} for $\Vec{v}_{\rm drift}$, we have
 \begin{align}
     \langle|\vec{v}_{\rm drift}|\rangle\sim \frac{v_A^2}{a\lambda_D X_i\alpha_{\rm in}},
 \end{align}
 where we used the fact that the characteristic length-scale on which $\Vec{B}$ changes is given by the damping scale $\lambda_{\rm D}$, and $v_A = a^{-4}\langle \Vec{B}^2\rangle /4\pi \rho_{\rm b}$ is the Alfvén velocity. Similarly, the average velocity of the whole fluid can be found by equating the l.h.s and r.h.s term in Eq.~\eqref{eq:Lor_for_neutrals} to yield
 \begin{align}
     \langle |\vec{v}_{\rm b}|\rangle\sim \left\langle  \frac{1}{Ha^5} \frac{\lvert(\nabla \times \Vec{B}) \times \Vec{B}\rvert}{4\pi \rho_{\rm b}}\right \rangle\sim \frac{v_A^2}{a\lambda_DH}.
 \end{align}
 Then by imposing $\langle|\vec{v}_{\rm drift}|\rangle \ll \langle |\vec{v}|\rangle$ and using the fact that $\lambda_{\rm D} \simeq v_{A}/aH $ (see Eqs.~\eqref{eq:lambdaD_va}, \eqref{eq:lambda_sim}, \eqref{eq:lamkap} in the main text), we obtain a condition for the ionization fraction
 \begin{equation} \label{eq:ionization_cond}
     X_{\rm i} \gg X_{\rm cr}\equiv\frac{H(z)}{\alpha_{\rm in}} = \frac{H_0 \sqrt{\Omega_{\rm m}} (1+z)^{3/2}}{\alpha_{\rm in}} \,.
 \end{equation}
%\begin{equation}
%   \frac{\langle \lvert (\nabla \times \Vec{B})\times \Vec{B} \rvert\rangle }{4\pi  a^5\rho_{\rm i}\alpha_{\rm in} } \simeq \frac{v_A^2}{\lambda_{\rm D}} \ll a X_{\rm i} \alpha_{\rm in} <|v|>  \,,
%\end{equation}
To find the value of $X_{\rm cr}$, we need to first quantify $\alpha_{\rm in}$. After recombination, and before the first stars begin to form, the most abundant neutral atoms are hydrogen (74\%) and helium (25\%), with traces of molecular hydrogen $\rm H_2$ and metals \cite{PeeblesCosmology,ParticleDataGroup:2022pth}. In the ion-neutral interaction rate, the dominant contribution  will come from proton-hydrogen collisions. Thus, we take \cite{Draine2011}
\begin{equation}
    \langle \sigma w\rangle_{\rm in} = \langle \sigma w\rangle_{\rm H^{+} H} = 3.25\times 10^{-9} {\rm cm^3 s^{-1}} \,,
\end{equation}
and we will use $\alpha_{\rm in} = n_{\rm H} \langle \sigma w\rangle_{\rm H^{+} H} /2$. We can write approximately $n_{\rm H} \approx \rho_{\rm b}/(1.3 m_{\rm H}) \simeq 8.6 \times 10^{-6} \Omega_{b} h^2 (1+z)^3 {\rm cm^{-3}}$.
Substituting the above relation in Eq.~\eqref{eq:ionization_cond}, we obtain
\begin{equation} \label{eq:Xi_ineq}
  X_{\rm cr} =\frac{2.3 \times 10^{-4}}{(1+z)^{3/2}} \frac{\sqrt{\Omega_{\rm m}}}{\Omega_{\rm b}h}  \,.
\end{equation}
From the theory of recombination, one has the following formula for the residual ionization fraction after redshift $z\sim 200$, fitted at $z=100$ \cite{PeeblesCosmology}
\begin{equation}
    X_{\rm i} \simeq 1.2 \times 10^{-5} \frac{\sqrt{\Omega_{\rm m}}}{\Omega_{\rm b} h} \,.
\end{equation}
Thus we have 
\begin{equation}
    \frac{X_{\rm i}}{X_{\rm cr}} \approx 50 \left(\frac{1+z}{100}\right)^{3/2}.
\end{equation}
One can see that for $z \sim 100$ the inequality in Eq.~\eqref{eq:ionization_cond} is satisfied by at least an order of magnitude. For $z<100$, as discussed in the main text, gravity overcomes the Lorentz force. As gravity acts equally on ions and neutrals, tight coupling approximation is irrelevant in the context of structure formation for $z<100$.

The breakdown of tight-coupling approximation is closely related to the ambipolar Reynolds number discussed in earlier literature \cite{Banerjee:2004df}. Specifically, the drift between ions and neutrals introduces an additional dissipation mechanism for the magnetic fields, which can be seen by replacing $\Vec{v}_{\rm i} \approx \Vec{v}_{\rm b} + \Vec{v}_{\rm drift}$ into \eqref{eq:AD_induction},
\begin{equation}   \label{eq:ind_amb}
    \frac{\partial \Vec{B}}{\partial t} = \frac{1}{a}\nabla \times (\Vec{v}_{\rm b} \times \Vec{B}) + \frac{1}{a^6}\nabla \times \left( 
    \frac{(\nabla \times \Vec{B})\times \Vec{B}}{4\pi \rho_{\rm i} \alpha_{\rm in}}  \times \Vec{B} 
\right)\,,
\end{equation}
where we substituted for the drift velocity from \eqref{eq:drift_vel}. The first term on the r.h.s. describes the standard advection (cf.~\eqref{eq:induction}), and the second term accounts for the ambipolar diffusion.

The ambipolar Reynolds number is defined by comparing the average of the two terms on the r.h.s, 
\begin{align}
    {\rm Re}_{\rm amb}=\frac{\langle |\vec{v}_{\rm b}|\rangle}{\langle |\vec{v}_{\rm drift}|\rangle}.
\end{align}
It can be seen that requiring ambipolar diffusion to be unimportant, i.e. ${\rm Re}_{\rm amb}\gg 1$, is the same as the tight coupling condition. One should bear in mind that even for ${\rm Re}_{\rm amb}\gg 1$, ambipolar diffusion can still heat the gas and alter the ionization history. This effect is strong at very small scales, below $\lambda_{\rm D}$ (for details, see \cite{Sethi:2004pe}).

	\section{Generating initial conditions with PMFs}\label{sec:ngenic}
	In this section, we outline the algorithm behind our code that generates initial conditions for baryons and dark matter for hydrodynamical simulations. We have modified the publicly available code, {\scshape N-GenIC}, to produce initial conditions with PMFs.
	
	\subsection{Brief review of the {\scshape N-GenIC} algorithm}
	The {\scshape N-GenIC} code takes in as input the density power spectrum of baryons and dark matter at a desired high redshift. Then as output, it produces the position of dark matter and baryon particles on a grid and also assigns each particle a velocity vector.
	
	The {\scshape N-GenIC} code utilizes the Zeldovich approximation to obtain displacement fields from density perturbations. Specifically, in the linear limit, the continuity equation is of the form
	\begin{align}
		\frac{d\delta}{dt}=-\frac{\nabla\cdot \vec{v}}{a}.
	\end{align}
	Integrating the above yields,
	\begin{align}
		\delta=-\nabla\cdot\vec{x}_{\rm dis}.
	\end{align}
	Here $\vec{x}_{\rm dis}$ is the displacement vector from original position. The original position should be such that $\delta=0$, i.e. when particles are distributed uniformly on a grid.
	
	Finding the displacement vector field, $\vec{x}_{\rm dis}$, from density perturbations, $\delta$, is simplified in Fourier space, where
	\begin{align}
		\vec{x}_{\rm dis}(k)=-\frac{\vec{k}}{k^2}\delta(k).
	\end{align}
	In the above equation, a critical assumption has been made that particles are not displaced in directions perpendicular to $\vec{k}$, i.e. there is no vortical motion. This assumption is appropriate in $\Lambda$CDM cosmology as there are no sources of vortical motions. 
	
	Thus, {\scshape N-GenIC} first initializes the displacement vector field in Fourier space and then takes the inverse Fourier transform to obtain the displacements in real space. The velocity vector for each particle is obtained by simply multiplying their displacement vector with the Hubble rate, $H$.

	\subsection{{\scshape N-GenIC} with PMFs}
	In this work, we modified {\scshape N-GenIC} to produce initial particle positions from a given magnetic field power spectrum. We do so by first generating vector potential $A$ on our grid, where $\vec{B}=\nabla\times \vec{A}$. Then we calculate $\vec{B}$ and displace particles according to the Lorentz force, $\vec{L}_{\rm B}=(\nabla\times\vec{B})\times\vec{B}$.
	
	\subsubsection{Initializing $\vec{B}$ fields}
	The power spectrum of $\vec{A}$ is related to that of magnetic fields as
	\begin{align}
		\epsilon_{\rm irs}\epsilon_{\rm jab}k_{\rm r}k_{\rm a}\langle A_{\rm b}(k) A_{\rm b}(k')\rangle=(2\pi)^3\delta^3(k-k')\left(\delta_{\rm ij}-\frac{k_{\rm i}k_{\rm j}}{k^2}\right)\frac{P_{\rm B}(k)}{2}.
	\end{align}
	In the above form it is hard to understand the spectrum of $A$ fields. Things are simplified if we change our coordinates, such that the $z$-axis is along $\vec{k}$. In this coordinate, we obtain
	\begin{align}
		\langle A_2 A_2\rangle=\langle A_1 A_1\rangle=(2\pi)^3\delta^3(k-k')\frac{P_{\rm B}(k)}{2k^2}, && \langle A_1 A_2\rangle=0. 
	\end{align}
	here the subscript $1$ and $2$ are simply two directions perpendicular to $\vec{k}$. The value of $A_3$ does not contribute towards the final value of the B field and hence we set it to zero.
	
	At every grid point in the Fourier space, we first initialize the value of vector fields ($A_1$ and $A_2$) in the directions perpendicular to $\vec{k}$. Considering direction of $\vec{k}$ is of form
	\begin{align}
		\hat{k}=(\sin\theta\sin\phi,\sin\theta\cos\phi,\cos\theta),
	\end{align}
	and choosing one perpendicular direction to $\vec{k}$ to lie in $\hat{k}-\hat{z}$ plane and the other to lie in $x-y$ plane, we obtain
	\begin{align}
		A_z=A_1\sin\theta && A_x=A_2\cos\phi-A_1\cos\theta\sin\phi && A_y=-A_2\sin\phi-A_1\cos\theta\cos\phi.
	\end{align}
	
	The above procedure gives us $\vec{A}$ field in Fourier space. We then inverse Fourier transform $\vec{A}$ to obtain its values in real space. Subsequently, the magnetic fields are obtained by evaluating $\nabla\times \vec{A}$ in the real space. This procedure ensures that the magnetic fields are generated with minimal divergence.
	
	\begin{figure}
		\begin{subfigure}{0.5\textwidth}
			\includegraphics[width=1.00\textwidth]{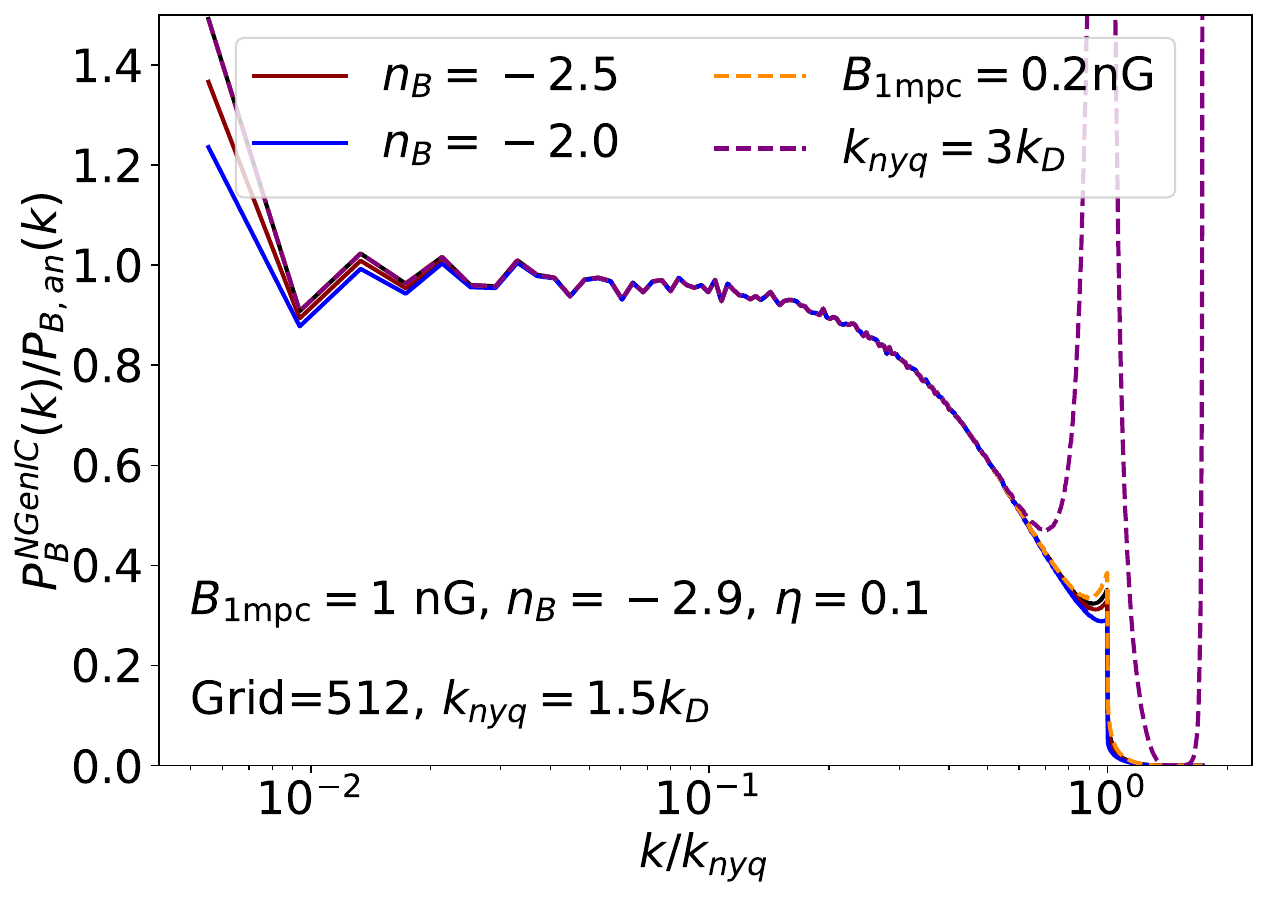}
		\end{subfigure}
		\begin{subfigure}{0.5\textwidth}
			\includegraphics[width=1.00\textwidth]{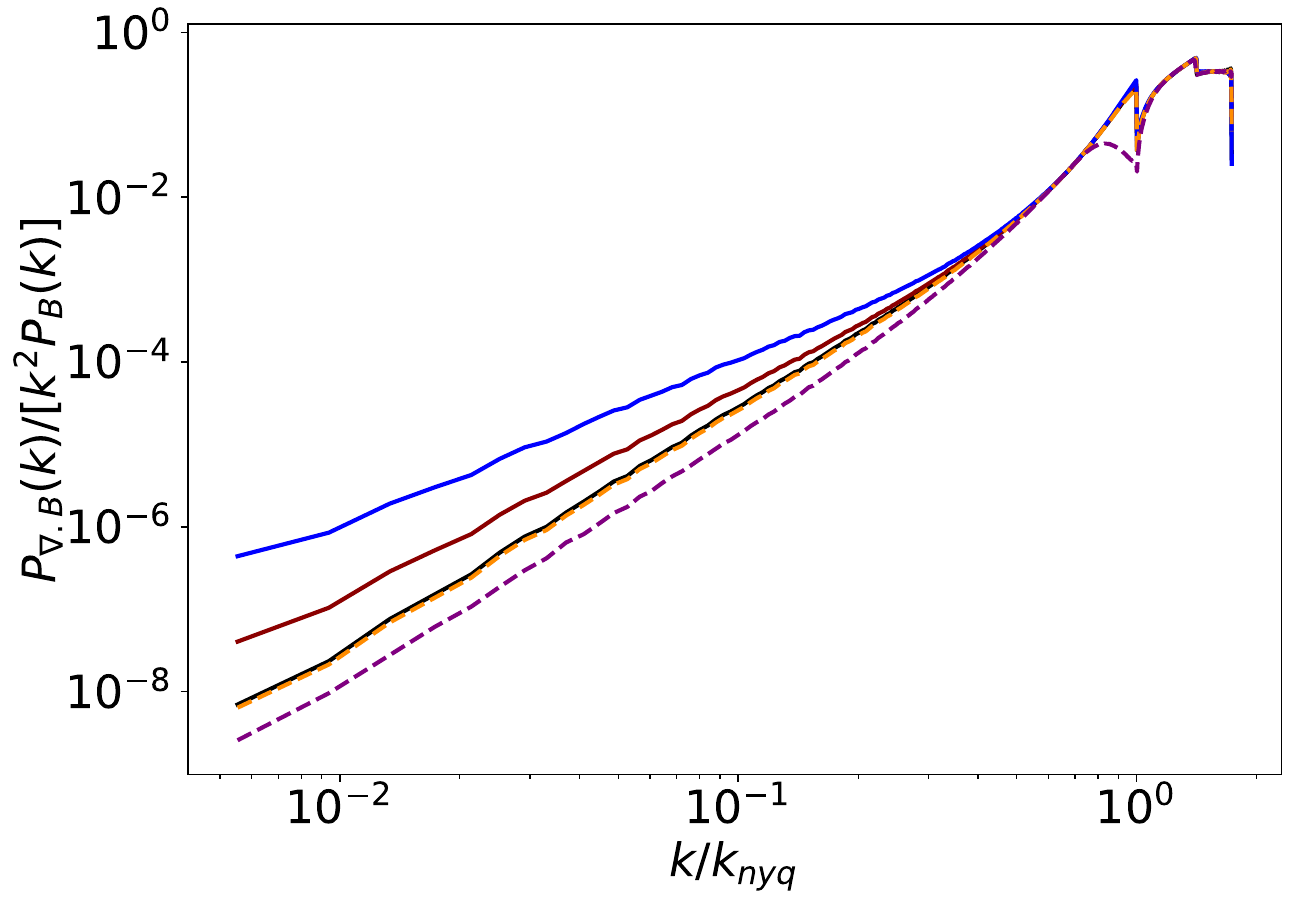}
		\end{subfigure}
		
		\caption{\textbf{Left:} The ratio of the magnetic field power spectrum numerically obtained from {\scshape N-GenIC} compared to the input power spectrum in the {\scshape N-GenIC} code. The black line is for the parameter point given in the text. The coloured lines are obtained after varying the parameters in the legend. This figure highlights that our modified {\scshape N-GenIC} code accurately initialized magnetic fields on the grid, with small deviations near Nyquist frequency, $k_{nyq}$. \textbf{Right:} The power spectrum of $\nabla\cdot \vec{B}$ compared with $k^2P_{\rm B}$. The small value of this ratio highlights that magnetic fields generated by {\scshape N-GenIC} have negligible divergence.}\label{fig:PB_ngenic_test_A}
	\end{figure}
	
	In the left panel of Fig.~\ref{fig:PB_ngenic_test_A} we show the power spectrum of magnetic fields generated by {\scshape N-GenIC} and compare it with the input power spectrum of magnetic fields. One can see that {\scshape N-GenIC} accurately produces the magnetic fields in the desired way. The right panel of Fig.~\ref{fig:PB_ngenic_test_A} shows that the divergence of magnetic fields generated by {\scshape N-GenIC} is numerically insignificant.
	
	\subsubsection{Initializing particle positions}
	To obtain the displacement vector for baryons due to magnetic fields, we integrate the baryon Euler equation (Eq.~\eqref{eq:vb}) and neglect the contribution from gravity to obtain
	\begin{align}\label{eq:dis_PMF}
		\vec{x}_{\rm dis}^{\rm PMF}\propto \frac{(\nabla\times \vec{B})\times \vec{B}}{(4\pi\rho_ba^3)a^3H^2}.
	\end{align}
	Note that $\vec{x}_{\rm dis}^{\rm PMF}$ vector field has both irrotational and vortical (divergence-free) components in it, unlike the displacements generated in $\Lambda$CDM cosmology.
	
	Considering the dimensionless proportionality constant in Eq.~\eqref{eq:dis_PMF} to be $\xi^{\rm num}$, the total displacement for a particle is simply given by adding the displacement from $\Lambda$CDM power spectrum with the displacement from PMFs,
	\begin{align}\label{eq:dis_full}
		\vec{x}_{\rm dis}=[\vec{x}_{\rm dis}]_{\rm \Lambda CDM}+\xi^{\rm num}\times \frac{(\nabla\times \vec{B})\times \vec{B}}{(4\pi\rho_ba^3)a^3H^2}.
	\end{align}
	The above prescription ensures that when perturbations are in the linear regime, the density perturbation field is given by
	\begin{align}\label{eq:zeldovich_pmf}
		\delta=-\nabla\cdot\vec{x}_{\rm dis}=\delta_{\rm \Lambda CDM}-\xi^{\rm num}\frac{S_0}{a^3H^2},
	\end{align}
	which matches what we expect from theory, see Eq.~\eqref{eq:xi_def}.

    \subsection{Shortcomings of our initial conditions}
    Our current algorithm for producing initial conditions has three shortcomings.
    
    The most straightforward and perhaps unimportant shortcoming in our initial conditions is that the prescription for displacement given in Eq.~\eqref{eq:dis_full} would also produce a vortical displacement for dark matter. However, as dark matter only responds gravitationally to baryons, they should only have motion in irrotational directions. Since the vortical motion in matter particles decays as $1/a^2$ and as dark matter motion is suppressed compared to baryons by a factor of 10 at $z=99$, we expect negligible effect from the artificial vortical motion in dark matter.

    The second limitation is that our procedure cannot generate appropriate initial conditions for scales near and beyond the damping scale, $\lambda_D$. This is because, on scales near the damping scale, the baryon perturbations reach non-linear values and the Zel\'dovich approximation (Eq.~\eqref{eq:zeldovich_pmf}) breaks down. Thus, we choose our grid and box sizes such that Nyquist frequency is near $k_D=1/\lambda_D$.
    
    Finally, the most important limitation is that the power spectrum obtained from our initial conditions is suppressed compared to the analytical estimate if we set $\xi^{\rm num}$ in Eq.~\eqref{eq:zeldovich_pmf} to their theoretically derived values in Eq.~\eqref{eq:xi_def}. Our numerical power spectrum is suppressed for two reasons.
	
	First, because the hydrodynamical simulations do not include MHD and hence do not capture the support provided by the $S_0$ term (see Eq.~\eqref{eq:delta_b}) to the growth of density perturbations after $a>0.01$. While gravity does quickly dominate over the $S_0$ term by $a\gtrsim 0.01$, the takeover is gradual. Thus the hydrodynamical simulation would slightly underestimate the final density perturbation. To counter this underestimation, we need to slightly increase the value of $\xi$ provided to the Gadget simulation to obtain the correct value of density perturbations at late times.
	
	\begin{figure}
		\centering
		\includegraphics[width=0.7\textwidth]{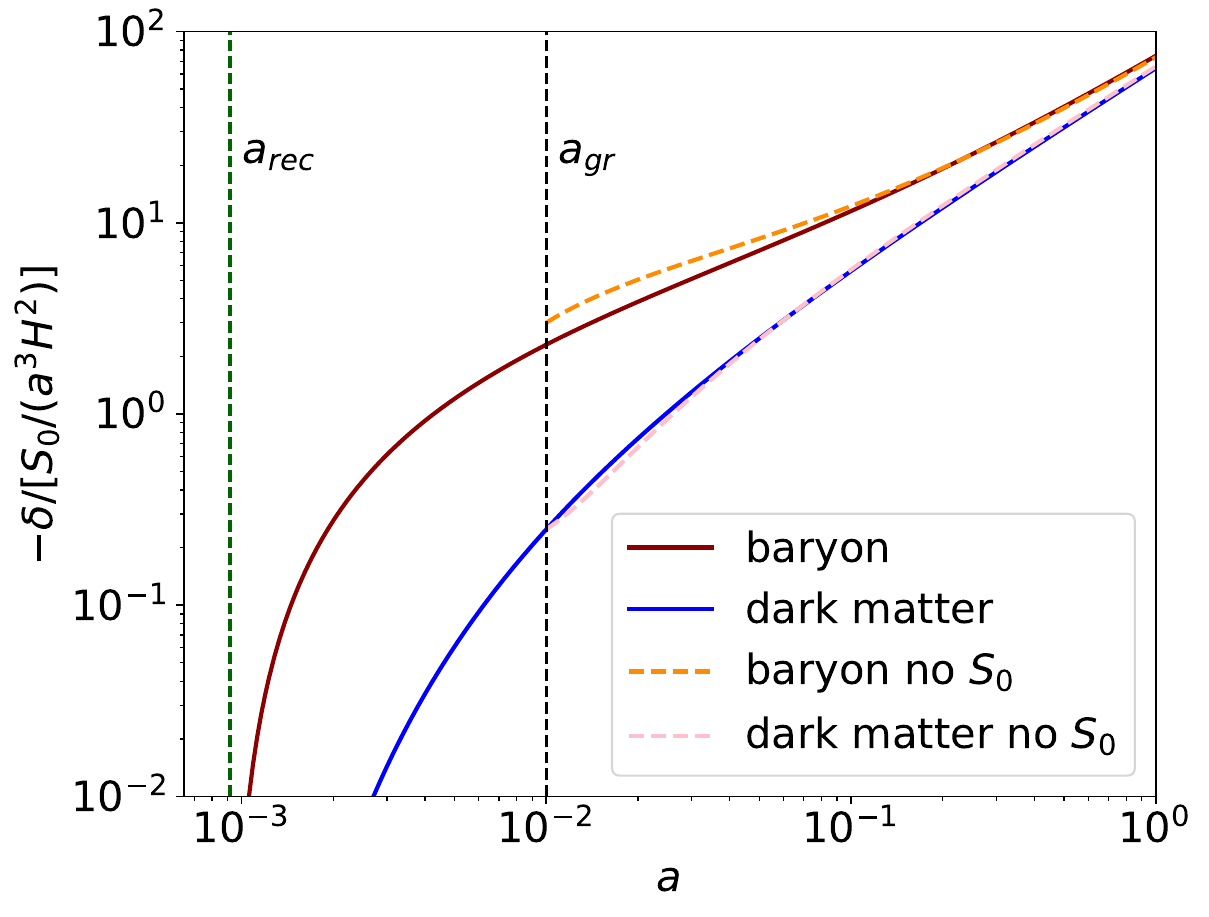}
		
		\caption{Evolution of density perturbations that are sourced by PMFs and normalized with $S_0(k)/[a^3H^2]$. The dashed lines are solutions with boosted initial conditions but with $S_0$ set to zero in the evolution equation. The solid line shows the real solution that includes $S_0$.}\label{fig:lambda_check}
	\end{figure}
	
	In Fig.~\ref{fig:lambda_check}, we show the evolution of $\delta$ normalized to $S_0/[a^3H^2]$ after solving the pertubation equations given in Eqs.~\eqref{eq:deltab_a}-~\eqref{eq:deltadm_a} as solid lines. The dashed lines show the evolution when $S_0$ term is set to zero in Eqs.~\eqref{eq:deltab_a} and the initial condition for $\delta_{\rm b}$ and $\delta_{\rm DM}$ are taken at $a=0.01$ with $\delta_{\rm b}$ initial value increased by 30\%. The initial condition for ${\rm d}\delta/{\rm d}a$ is simply taken as ${\rm d}\delta/{\rm d}a=\delta/a$, which is effectively the same as the initial condition employed by {\scshape N-GenIC} for velocities. We can see that by marginally boosting the initial condition, the gravity-only equation can correctly obtain the same late-time behaviour as the equation with both gravity and Lorentz force.
	
	\begin{figure}
		\begin{subfigure}{0.5\textwidth}
			\includegraphics[width=1.00\textwidth]{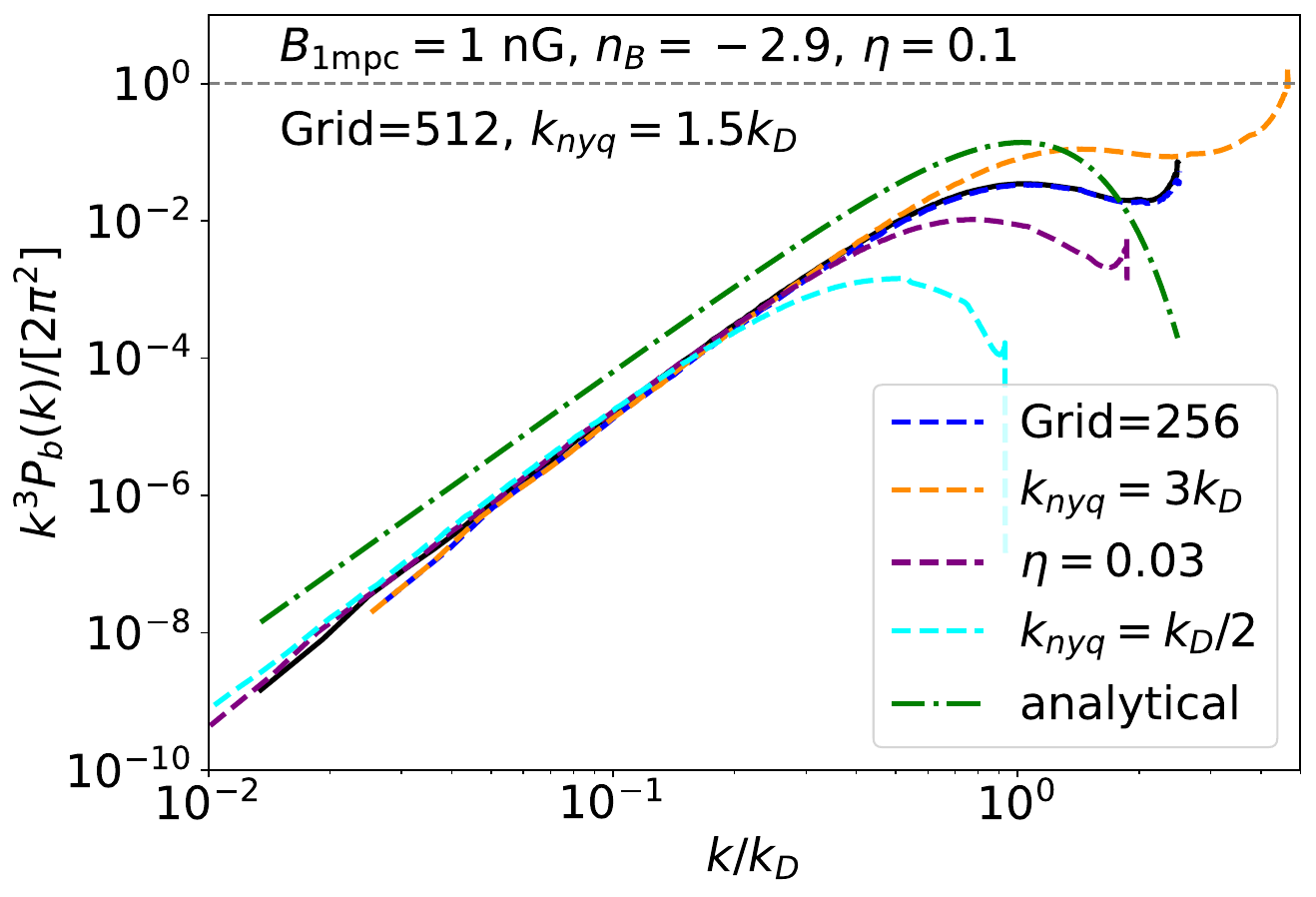}
		\end{subfigure}
		\begin{subfigure}{0.5\textwidth}
			\includegraphics[width=1.00\textwidth]{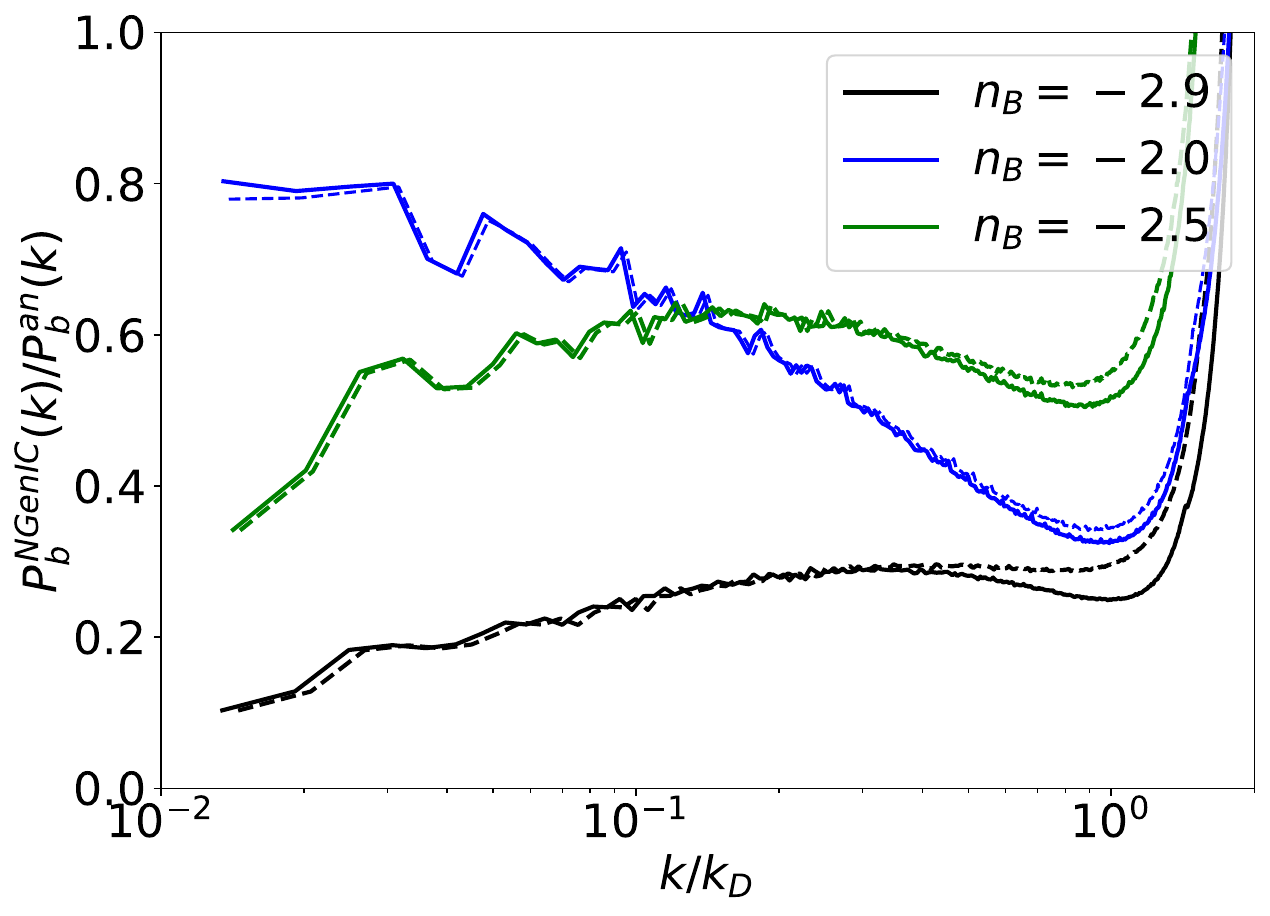}
		\end{subfigure}
		
		\caption{\textbf{Left:} Dimensionless baryon power spectrum generated from PMFs at $z=100$ after turning off contribution from inflationary initial conditions. The black line is the power spectrum generated by {\scshape N-GenIC} for parameters given in the text. The dashed lines correspond to the power spectrum also obtained by {\scshape N-GenIC} but after varying the parameters mentioned in the legend. The green dashed line shows the analytical estimate obtained using Eq.~\eqref{eq:Delta_b} but with $\xi_{\rm b}(0.01)$ enhanced by 30\% to match with the $\xi_{\rm b}$ set in N-GenIC. This figure highlights that our numerical code is self-consistent and the deviation from the analytical estimate is not due to resolution issues. \textbf{Right:} The ratio of numerical baryon power spectrum with the analytical expectation for different values of $n_{\rm B}$ and $B$ but with $\eta=0.1$. The solid lines are for $B_{\rm 1Mpc}=1$ nG and the dashed lines are for $0.2$ nG. For all the lines, the grid is divided into $512$ parts and the box size is set such that $k_{nyq}=1.5k_{\rm D}$. This figure illustrates that the deviation from analytical estimate is only sensitive to $n_{\rm B}$.}\label{fig:PS0_ngenic_test_A}
	\end{figure}
	
	The second reason for the suppressed value of our numerical power spectrum is because the power spectrum of the $S_0$ field produced by {\scshape N-GenIC} is smaller than what is analytically expected. In the left panel of Fig.~\ref{fig:PS0_ngenic_test_A}, we show the baryon power spectrum calculated from {\scshape N-GenIC} after turning off contribution from inflationary initial conditions. We see that there is an amplitude mismatch of order $\sim$10 between the numerical and analytical estimate, and this mismatch is robust even with an increase in resolution. In the right panel, we show that the mismatch varies for different values of $n_B$ but is insensitive to the strength of magnetic fields, $B_{\rm 1 Mpc}$. The value of mismatch is the same also for dark matter.
	
	The reason for the above mismatch between the analytical expectation and the numerical result is not certain. We conjecture that it could be due to non-negligible divergence of $B$ near Nyquist frequencies propagating to smaller wavenumbers of $S_0$.
	
	To counter the numerical underestimation of our baryon and dark matter power spectrum, we choose an enhanced value of $\xi^{\rm num}$ such that the power spectra match the analytical expectation (see table~\ref{table:mod_params}).
	
	%Next, notice that the damping scale of $S_0$ obtained by NGenIC is also displaced to slightly larger wavenumbers than analytical expectation. To make the numerical damping scale match the analytical one, we use a smaller value of $\eta$ in our numerical code, which we call $\eta^{\rm num}$.
	
	\begin{table}[h]
		\centering
		\begin{tabular}{c c c c c } 
			\hline
			$n_{\rm B}$ & $\xi_{\rm b}^{\rm num}/\xi_{\rm b}(0.01)$ & $\xi_{\rm DM}^{\rm num}/\xi_{\rm DM}(0.01)$    \\ 
			\hline
			-2.9  & $1.3 \times 2.23$ & $2.23$\\
			-2.0 & $1.3 \times 1.20$ & 1.20\\
			\hline
		\end{tabular} \\
		\caption{{\scshape Modified Theory Parameters of simulations}. The first column describes the maximum peak value of the baryon power spectrum (parameterized by $\eta$) according to analytical estimate at $z=100$. The next two columns show the additional boost provided to baryon and dark matter particles respectively. The first number in the second column represents the boost needed to balance the absence of MHD in our simulation. The second number represents the boost needed to counter the suppressed value of $S_0$ power spectrum. For dark matter particles only the second boost is required.}
		\label{table:mod_params}
	\end{table}
	
	\subsection{Fixing $\eta$}\label{sec:eta_fix}
	As mentioned in section~\ref{sec:power}, we introduce an additional parameter $\eta$ (eq.~\eqref{eq:eta}) to limit baryon perturbations from obtaining large non-linear values in the initial conditions. In this subsection, we expand on our reasoning behind limiting $\eta<0.3$.
	
	\begin{figure}
		\centering
		\includegraphics[width=0.7\textwidth]{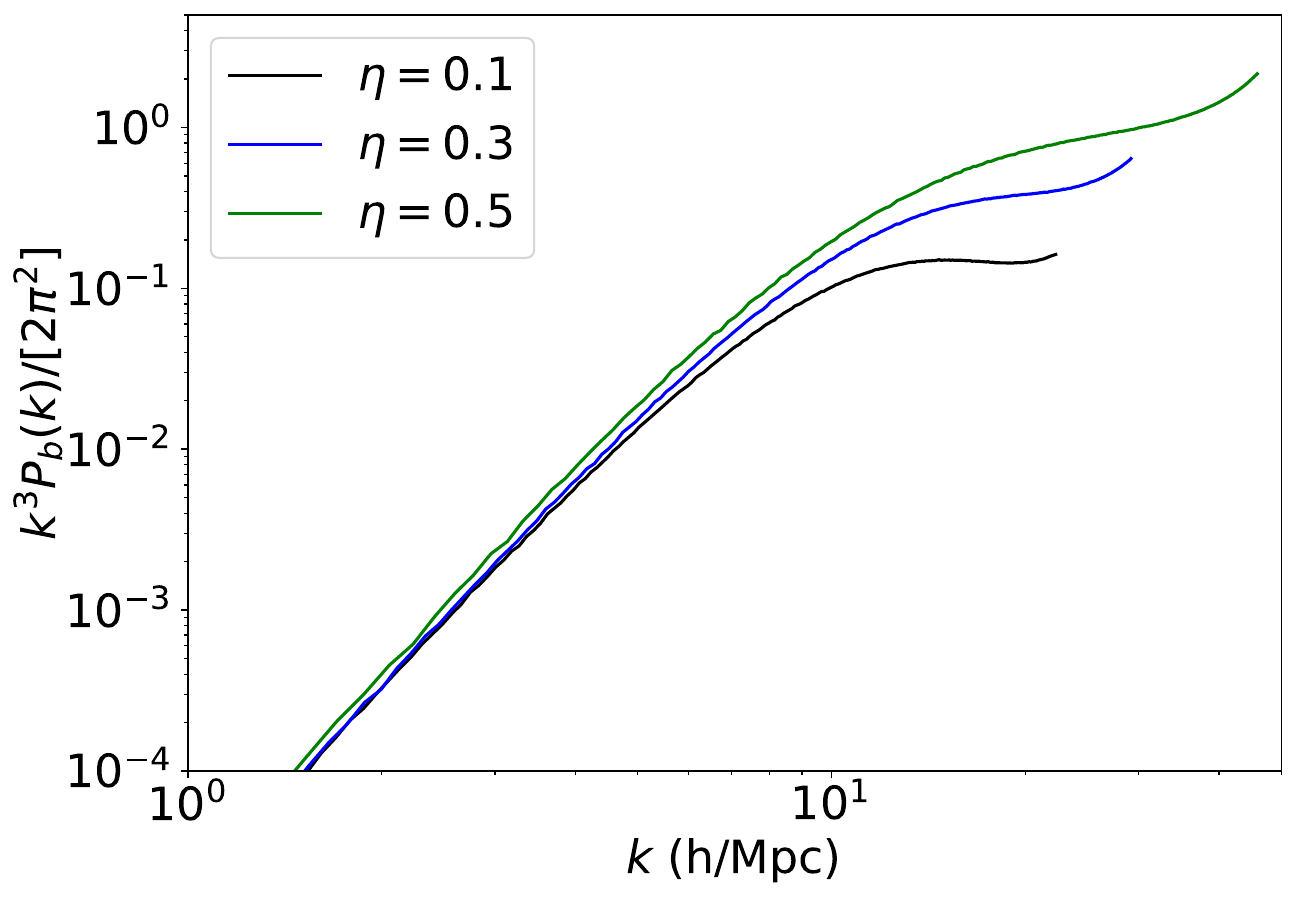}
		
		\caption{Dimensionless baryon power spectrum sourced by PMFs with $B_{\rm 1Mpc}=1$ nG and $n_B=-2.9$. The power spectra are shown for different values of $\eta$ and after turning off contribution from $\Lambda$CDM perturbations.}\label{fig:eta_check}
	\end{figure}
	
	In figure~\ref{fig:eta_check} we show how the PMF-induced power spectrum varies for different choices of $\eta$. Note that the initial conditions are generated with the $\xi^{\rm num}$ values as specified in table~\ref{table:mod_params}. As expected, larger $\eta$ values lead to larger peak in the baryon power spectrum. However, for $\eta\gtrsim 0.3$ we find that the baryon power spectrum has no discernable maximum and continues to increase until the Nyquist frequency. Moreover, as we increase $\eta$, the power spectrum at wave numbers much smaller than the Nyquist frequency increases. For instance, in figure~\ref{fig:eta_check} one can see that the power spectrum for $\eta=0.5$ is slightly larger than the power spectrum for $\eta=0.3$ and $\eta=0.1$. We conjecture that the impact on large wavenumber is due to backreaction from non-linearity at small scales.
	
	Thus, to avoid backreaction and to not have a growing $k^3P_b$ beyond $k_D$, we limit $\eta<0.3$.
	
	\section{Probability distribution function of density fields sourced by PMFs}\label{sec:pdf}
	In $\Lambda$CDM cosmology, the initial condition of the density perturbations is a stochastic variable with a Gaussian distribution. Thus, even as the density perturbations evolve, they retain their Gaussian distribution as long as the perturbations are small, i.e. in the linear limit. In contrast, PMFs source density perturbations such that
	\begin{align}
		\delta_{\rm b}\propto \nabla\cdot[(\nabla\times\vec{B})\times\vec{B}].
	\end{align}
	As we consider $B$ fields to have a Gaussian distribution and as $\delta_{\rm b}$ is non-linearly related to $B$ fields, $\delta_{\rm b}$ cannot be Gaussian distributed.
	
\begin{figure}		\includegraphics[width=1.00\textwidth]{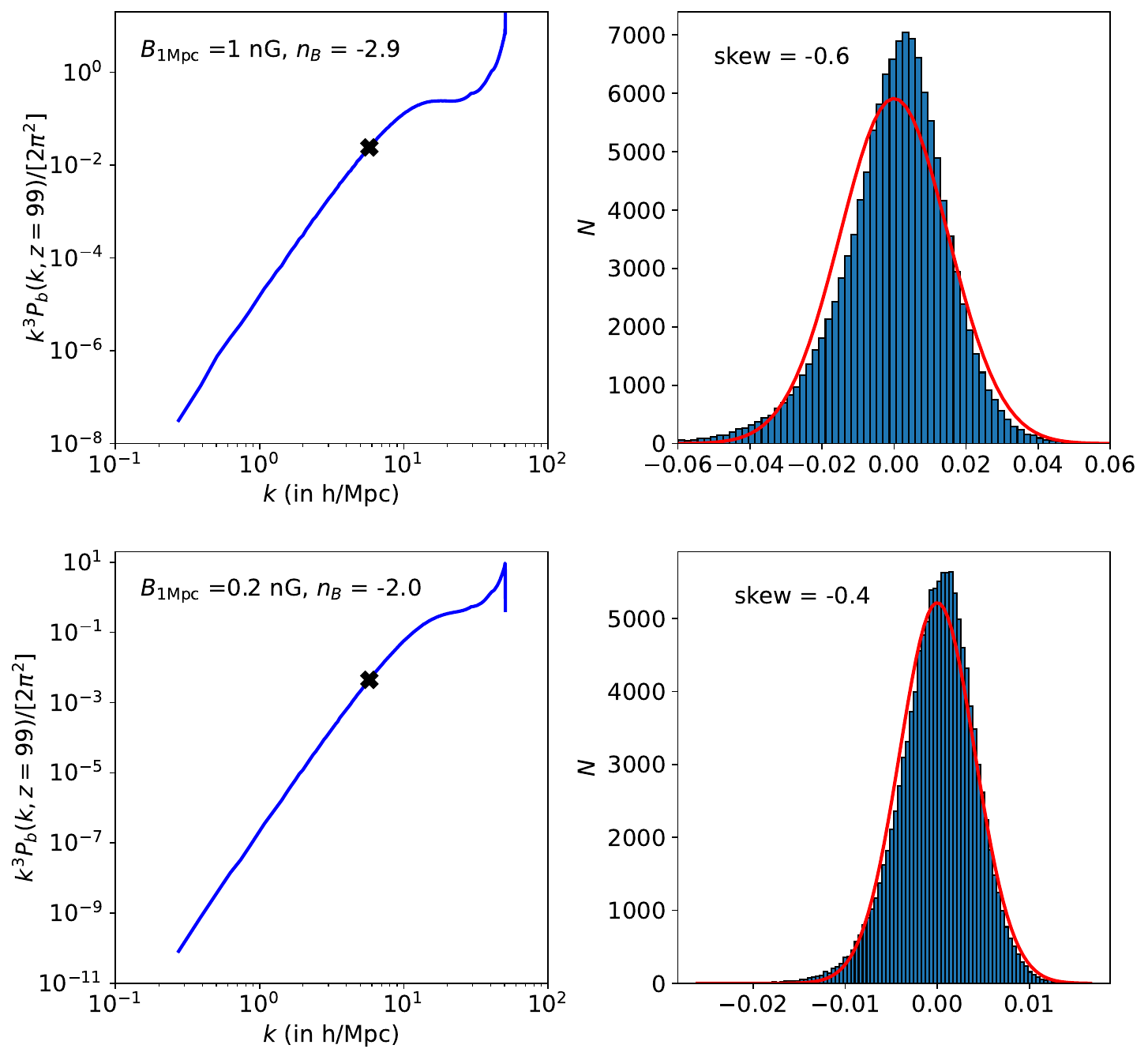}
		\caption{\textbf{Left} Baryon power spectrum of the initial conditions generated by our {\scshape N-GenIC} code. \textbf{Right:} Histogram of baryon perturbations smoothed on a scale marked by the black cross in the left panel. The red line shows the envelope of the histogram if the perturbations were Gaussian distributed with the same variance. }\label{fig:pdf}
	\end{figure}

	In Fig.~\ref{fig:pdf}, we show how density perturbations are distributed when they are sourced by PMFs and when  $\Lambda$CDM contribution is turned off. One can see that density perturbations have a small negative skew, whose value decreases as we increase $n_{\rm B}$. Regardless, to a good approximation, one can consider the density perturbation to be Gaussian distributed.
	
	\acknowledgments
        The authors thank Paola Santini, Chiara Feruglio, Roberta Tripodi, Enrico Garaldi, Milena Valentini, Takeshi Kobayashi, Aseem Paranjape, and Shiv Sethi for useful conversations. We also thank the referee for useful comments that helped improve our paper. PR acknowledges support from INFN TASP. MV is partly supported by the Fondazione ICSC National Recovery and Resilience Plan (PNRR), Project ID CN-00000013 “Italian Research Center on HighPerformance Computing, Big Data and Quantum Computing” funded by MUR — Next Generation EU. All the simulations presented in this work have been run on the Ulysses supercomputer at SISSA.

	\bibliographystyle{utphys}
	\bibliography{references}
	
\end{document}